\documentclass[aps,pra,twocolumn,nopacs,floatfix,superscriptaddress]{revtex4}
\usepackage{epsfig}
\usepackage{graphicx}
\usepackage{dcolumn}
\usepackage{amsthm}
\usepackage{comment}
\usepackage[utf8]{inputenc}
\usepackage{blindtext}
\usepackage{caption}
\usepackage[labelformat=simple]{subcaption}

\usepackage{morefloats}
\usepackage{amsmath}
\begin{document}
\title{Generating sustained coherence in a quantum memory for
retrieval at the times of quantum revival}
\author{Tavshabad Kaur}
\affiliation{Department of Physics, Guru Nanak Dev
University, Amritsar, Punjab-143005, India}
\author{Maninder Kaur}
\affiliation{Department of Physics, Guru Nanak Dev
University, Amritsar, Punjab-143005, India}
\author{Arvind}
\affiliation{Department of Physical Sciences, Indian
Institute of Science Education and Research (IISER) Mohali,
PO Manauli, Punjab, 140306, India}
\affiliation{Vice Chancellor, Punjabi University Patiala,
147002, Punjab, India}
\author{Bindiya Arora}
\affiliation{Department of Physics, Guru Nanak Dev
University, Amritsar, Punjab-143005, India}
%%%%%%%%%%%%%%%%%%%%%%%%%%%%%%%%%%%%%%%%%%%%%%%%%%
\begin{abstract}
We study the time degradation of quantum information stored
in a quantum memory device under a dissipative environment
in a parameter range which
is experimentally relevant. The quantum memory under consideration comprises of  an optomechanical
system with additional Kerr non-linearity in the optical
mode and an anharmonic mechanical oscillator with quadratic
non-linearity. 
Time degradation is monitored, both in terms of loss 
of coherence which is analyzed with the help of 
Wigner functions,  
as well as in terms of loss of amplitude of 
the original state studied as a function of time.
While our time trajectories
explore the degree to which the stored information degrades
depending upon the variation in values of  various
parameters involved,  we  suggest a set of
parameters for which the original information can be retrieved
without degradation. We come across a highly
attention seeking situation where the role played by the
non-linearity is insignificant and the system behaves as if
the information is stored in a linear medium. 
For this case, the
information retrieval is independent of the coherence revival
time and can be retrieved at any instant during
the time evolution.  
\end{abstract}
\maketitle
\section{Introduction}

Quantum memory is a device that stores quantum information
in the form of quantum states for later retrieval
~\cite{lvovsky2009nat,doi:10.1080/09500340.2013.856482}. Due
to an upsurge in the field of quantum computation and
quantum communication there has been an increasing demand of
devices that could store quantum information. However the
storage of quantum information is a challenging task due to
the presence of quantum decoherences effects. These effects
render the stored information unsuitable for further use.
The need of the hour is to devise 
 quantum memories that not
only can
preserve  coherence but 
also can store the
information for longer
durations. Criteria for assessing the
performance of quantum memory
include fidelity
~\cite{PhysRevLett.102.203601}, efficiency
~\cite{PhysRevLett.112.033601}, transfer coefficient
~\cite{Cao:20}, conditional variance
~\cite{PhysRevA.77.012323}, multimode capacity
~\cite{PhysRevA.79.052329,PhysRevLett.101.260502} and
storage time ~\cite{Vivoli_2013}.  Optomechanical systems
are highly recommended to store and retrieve quantum
states.  They provide a promising mechanism for a
high-fidelity quantum memory that is faithful to a given
temporal mode structure, and can be recovered synchronously.
Teh et al.~\cite{PhysRevA.96.013854} carried out quantum
simulations using a  non-linear optomechanical system to give
a complete model for the storage of a coherent state and to
increase the fidelity beyond the quantum threshold.
Recently, quantum memory protocol for the creation, storage
and retrieval of Schr\"odinger cat states in optomechanical
system was proposed by Teh et al.
~\cite{PhysRevA.98.063814}.  Quantum state formation through
system anharmonicities as well as decoherence due to
interactions with the environment in terms of Wigner current
was studied by Braasch et al.~\cite{PhysRevA.100.012124}.
Chakraborty et al.~\cite{Chakraborty:17} presented a scheme
to enhance the steady-state quantum correlations in an
optomechanical system by incorporating an additional
cross-Kerr-type coupling between the optical and the
mechanical modes.  He et al.~\cite{PhysRevA.79.022310}
proposed a dynamical approach for quantum memories using an
oscillator-cavity model to overcome the known difficulties
of achieving high quantum input-output fidelity with long storage
times.

Quantum memories are an indispensable part of quantum
information processing and long distance
communication
where the long distance is divided into shorter elementary
links and the quantum state is stored independently for each
link.  It finds numerous applications in quantum networks
~\cite{kimble2008quantum}, quantum repeaters
~\cite{PhysRevLett.81.5932,sangouard2011quantum} and linear
quantum computing ~\cite{knill2001scheme,RevModPhys.79.135}.
Quantum memory can be
used to enhance the
sensitivity of precision measurements where
the sensitivity is limited by the time over which phase can be
accumulated and hence storage of quantum states is required
~\cite{zaiser2016enhancing}.

In the present work we study an optomechanical system with
non-linearities to make it useful
for the purpose of a quantum memory.  The quantum
information is stored in the long-lived mechanical mode of the optomechanical system as a quantum state which
is initially chosen to be a coherent state obtained by
displacing the vacuum state by a certain 
amount~\cite{teh2018creation}.  The mechanical mode 
interacts with the optical mode  
through radiation pressure in such a
way that state transfer between both the modes is achievable
~\cite{palomaki2013coherent,reed2017faithful,PhysRevLett.108.153603}.  We
study the time evolution of our quantum state  under the
combined effect of non-linearity and dissipation using
the Master equation method.  The dissipation is modeled 
by assuming that the system interacts with a constant
temperature bath.  During the evolution highly
non-classical superpositions of coherent states, i.e.
multi-component Schr\"{o}dinger cat states (Kitten states)
are formed ~\cite{kirchmair2013observation}.  Revival of an
initial quantum state occurs when it evolves in time to a
state that reproduces its original coherent form. The
characteristic time scale over which this phenomenon happens
is called the coherence revival time
~\cite{rohith2015visualizing}.  The periodic revival
of coherence at the coherence revival times
makes the  retrieval of the quantum information at
these times possible.

We use the Wigner function in our analysis which
provides a particularly useful geometric representation of
quantum states as a real valued
function in the two-dimensional system
phase-space ~\cite{PhysRevA.100.012124,wigner1997quantum,hillery1984distribution,articleh,bookc}.
The time evolution of the Wigner function gives us
quantum phase-space dynamics.
We notice that the presence of non-linearity
causes the initial coherent state with associated
positive Wigner functions to evolve into non-classical
states  associated with negative Wigner functions
~\cite{PhysRevA.100.012124,Katz_2008}.  As expected
{the} coherent state resurrects itself 
maintaining its coherence at coherence revival time
{with certain amount of degradation due to the presence
of the environmental effects caused by the bath.}

Further, since quantum information is stored in the form of
a coherent state with certain amplitude that should be
maintained at coherence revival time, we
compute the amplitude of the coherent state
as a function of time and study the effect of decoherence
on the amplitude at different periods of time during
the evolution. On varying the values of bath temperature,
dissipation and non-linearity we gain insight into the
extent  to which coherences for
the evolved quantum state can be sustained.  In addition,
we propose the optimum parameters for optomechanical system
which can be used to retrieve quantum information at longer
times with minimum loss.

The paper is organized as follows: In Section~\ref{2} we present
the theory and the model used in our calculations, with the
Hamiltonian and relevant quantum states and their dynamics defend in
Section~\ref{2(a)}, the master equation and phase space
dynamics described in Section~\ref{2(b)} and the numerical
simulation techniques described in Section~\ref{2(c)}. In 
Section~\ref{3}  we discuss our results.  The
results are first presented in terms of Wigner function to
study the effect of decoherence on the quantum information
at different periods of time during evolution thereafter
temporal evolution of the amplitude is presented as function
of dissipation, non-linearity bath temperature and initial amplitude.
Section~\ref{4} contains some concluding remarks.

\section{Quantum Memory Model} \label{2}
A typical quantum optomechanical system consists of a
Fabry-Perot cavity with one of the movable mirrors acting as
a mechanical oscillator ~\cite{PhysRevA.96.013854}.  The optical
mode trapped inside the Fabry-Perot cavity is coupled  to the
mechanical mode via a generic coupling represented by
$g_0$ ~\cite{Chakraborty:17}.  The optical mode
and the mechanical mode have the frequencies
$\omega_{c}$ and
$\omega_{m}$, respectively.  
Additionally,
Kerr non-linearity of strength $k_c$ is present in the optical mode and $k_m$
being the strength of the quadratic non-linearity in the anharmonic
mechanical oscillator.
Practically, the optomechanical
system always interacts with its environment resulting in
decoherence through damping and fluctuations
~\cite{PhysRevA.96.013854}.  While interacting with their
corresponding reservoirs, the cavity decay rate is
represented by $\gamma_{c}$ and the
mechanical damping rate is given by
$\gamma_{m}$. As a result of such
dissipation, the quantum information 
begins to lose its
quality during the time evolution.

\begin{figure} \label{fig:Figure 1}
\includegraphics[height=5cm,width=9cm]{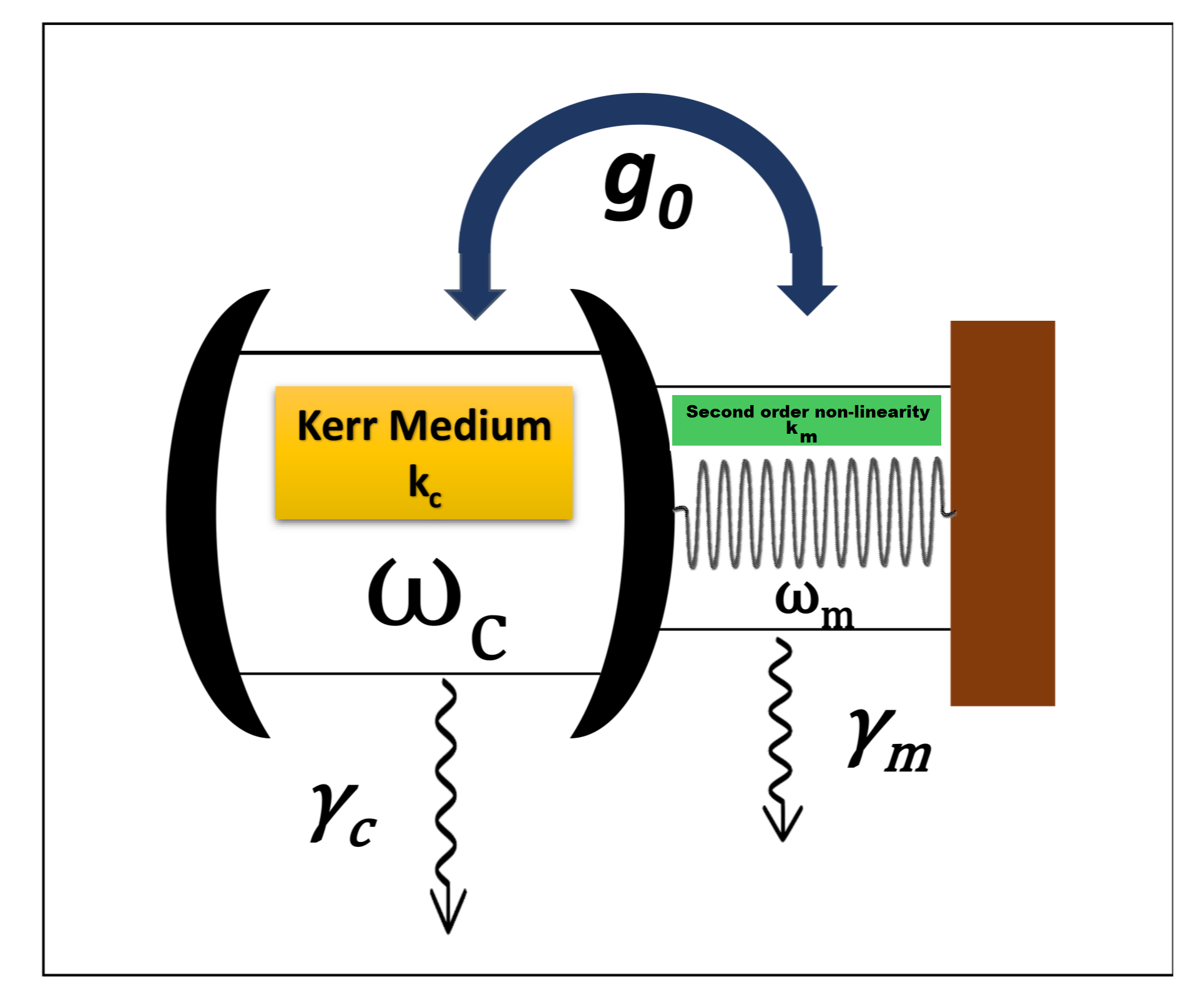}
\caption{Schematic diagram of optomechanical system taken as
quantum memory model. An optical mode (with frequency $\omega_{c}$ ) couples with
the mechanical mode (with frequency $\omega_{m}$ ) via the
radiation-pressure coupling $g_0$.  Additionally, Kerr
non-linearity of strength $k_c$ and quadratic non-linearity
of strength $k_m$ are included in optical and mechanical
modes, respectively.  Here, $\gamma_{c}$ is the optical decay
rate and $\gamma_{m}$ is the mechanical damping rate of the
respective modes.}
\end{figure}
%%%%%%%%%%%%%
\subsection{Hamiltonian for Quantum Information dynamics}
\label{2(a)} 
%%%%%%%%%%%%%%%
With {the} above considerations, the total Hamiltonian of the
optomechanical system 
can be written as 
\begin{equation} \label{eq:1}
H_{total}= H_{c}+H_{m}+H_{coupling}
\end{equation} 
where $H_{c}$, $H_{m}$ and $H_{coupling}$  
are 
{the Hamiltonian corresponding to the cavity, the
mechanical oscillator and the coupling between them,
respectively and are}
given by the following expressions:
\begin{eqnarray} 
\label{eq:2}
H_c&=&\hbar\omega_{c}a^{\dagger}a + \hbar
k_c(a^{\dagger}a)^2 \nonumber \\ 
H_m&=&\hbar\omega_{m}b^{\dagger}b + \hbar
k_m(b^{\dagger}b)^2\nonumber \\ 
H_{coupling}&=&-\hbar g_0 a^{\dagger}a(b+b^{\dagger}).
\end{eqnarray}
Here $a$($a^{\dagger}$) and $b$($b^{\dagger}$) are the
annihilation(creation) operators for optical
and mechanical modes, respectively.  {The terms}
$\hbar\omega_{c}a^{\dagger}a$  and $\hbar\omega_{m}b^{\dagger}b$
 correspond to the energy of the optical cavity
with frequency $\omega_{c}$ and mechanical oscillator with
frequency $\omega_{m}$ ~\cite{PhysRevA.96.013854}.
$H_{coupling}$ describes the interaction between the optical
mode and the mechanical mode  while
$\hbar k_c(a^{\dagger}a)^2$ and  $\hbar k_m(b^{\dagger}b)^2$ are the
non-linear terms present in the optical mode 
and the mechanical
oscillator, respectively.

In quantum information processing, quantum states are
prepared, stored and retrieved on-demand using certain
protocols. A unitary displacement operator $\mathcal{D}$
generates a coherent state $|\alpha\rangle$  by a
phase-space displacement of the vacuum state $|0\rangle$
~\cite{PhysRev.130.2529,PhysRevLett.10.277} as 
\begin{equation} \label{eq:4}
|\alpha\rangle = \mathcal{D}(\alpha)|0\rangle,
\end{equation}
where $\alpha$ is a complex parameter and  the Displacement operator  is given by
\begin{equation} \label{eq:6}
\mathcal{D}(\alpha)=exp (\alpha a^{\dagger}-\alpha^{*}a)
\end{equation} 
and $|\alpha\rangle$   can be
expanded in terms Fock or number states 
as~\cite{PhysRev.177.1857,articleh}:
\begin{equation} \label{eq:5}
|\alpha\rangle=e^{-\frac{|\alpha|^{2}}{2}}
\sum_{n=0}^{\infty}\frac{\alpha^{n}}{\sqrt{n!}}|n\rangle,
\end{equation}
This initially prepared  quantum coherent state is then
allowed to evolve under 
system-environment dynamics  to study the affect 
of dissipation and non-linearity on the stored information. 
For non-linear oscillators, after few cycles quantum 
interference takes 
over leading to a significant spread of the coherent state. 
The original state is no longer recognizable and is said to
 have ``collapsed".
As time advances, it resurrects itself leading to its ``revival". 
The time at which this revival takes place and initial
coherent state also regains its coherence is referred to as 
`coherence revival time' $T_{rev}$ ~\cite{kaur2018effect}. 
In optomechanical system we have two modes 
each with non-linearity such that both the modes behave as 
non- linear harmonic oscillators coupled with one another. 
Instead of considering these non-linear modes individually, 
{we work with the combined system as a whole}. 
{Therefore, the collapses and revivals occur for the two mode
system as a whole~\cite{kaur2018effect}. }
Consequently, we define  coherence revival time for our
system as  a whole
\begin{equation} \label{eq:7}
T_{rev}=\frac{2\pi}{k_c+k_m}.
\end{equation}
The entire numerical simulations under system-environment
dynamics are performed to bring forth the retrieval
phenomenon of stored quantum information at coherence
revival time {in the presence of environmental interactions.}

%%%%%%%%%%%%%%%%%%%
\begin{figure*}[ht]  
\centering
\begin{subfigure}{0.32\textwidth}
\includegraphics[height=4cm,width=4cm]{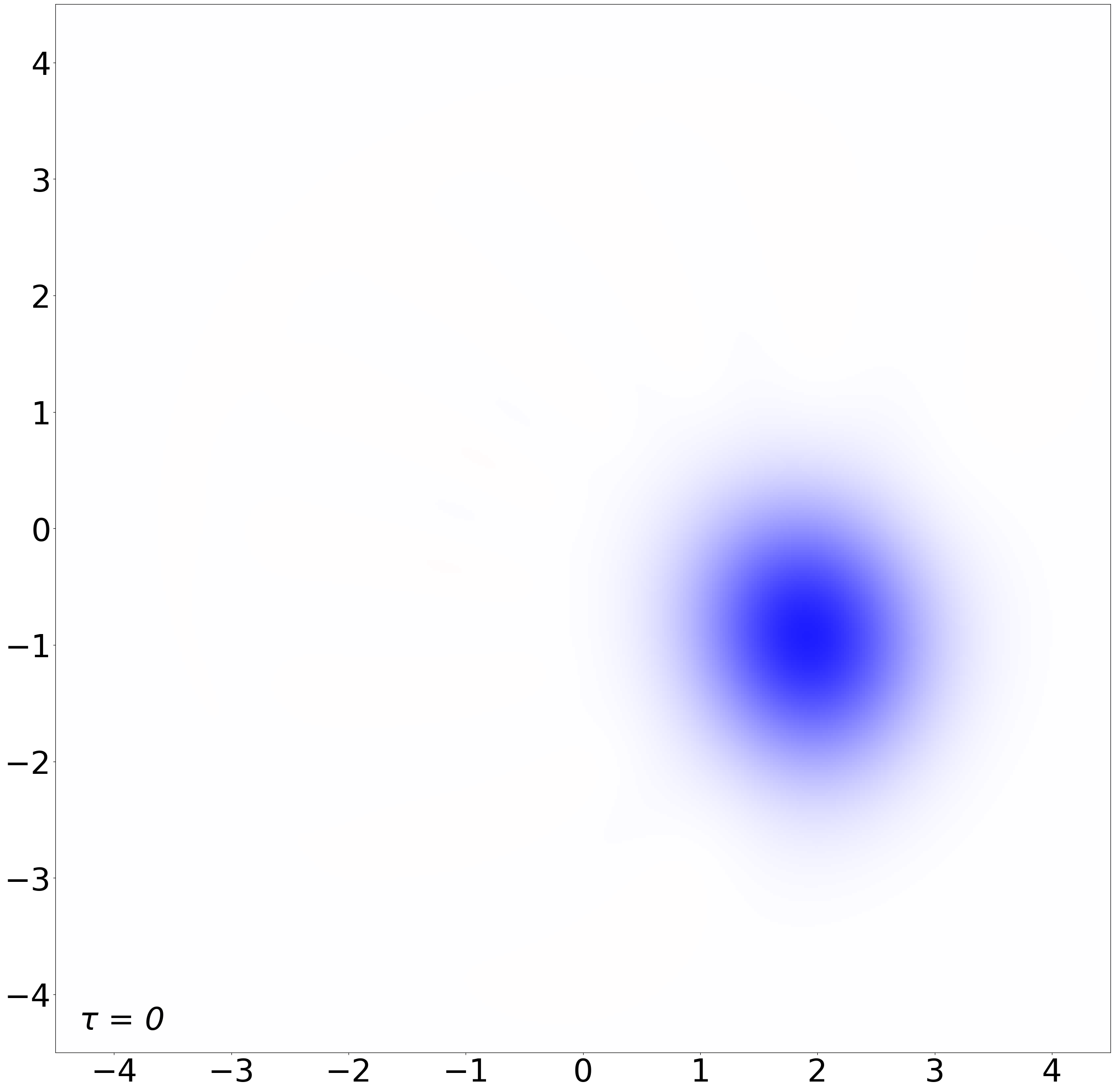}
\caption{$t=0$ (initial)}
\label{fig: 2(a)}
\end{subfigure}
\begin{subfigure}{0.32\textwidth}
\includegraphics[height=4cm,width=4cm]{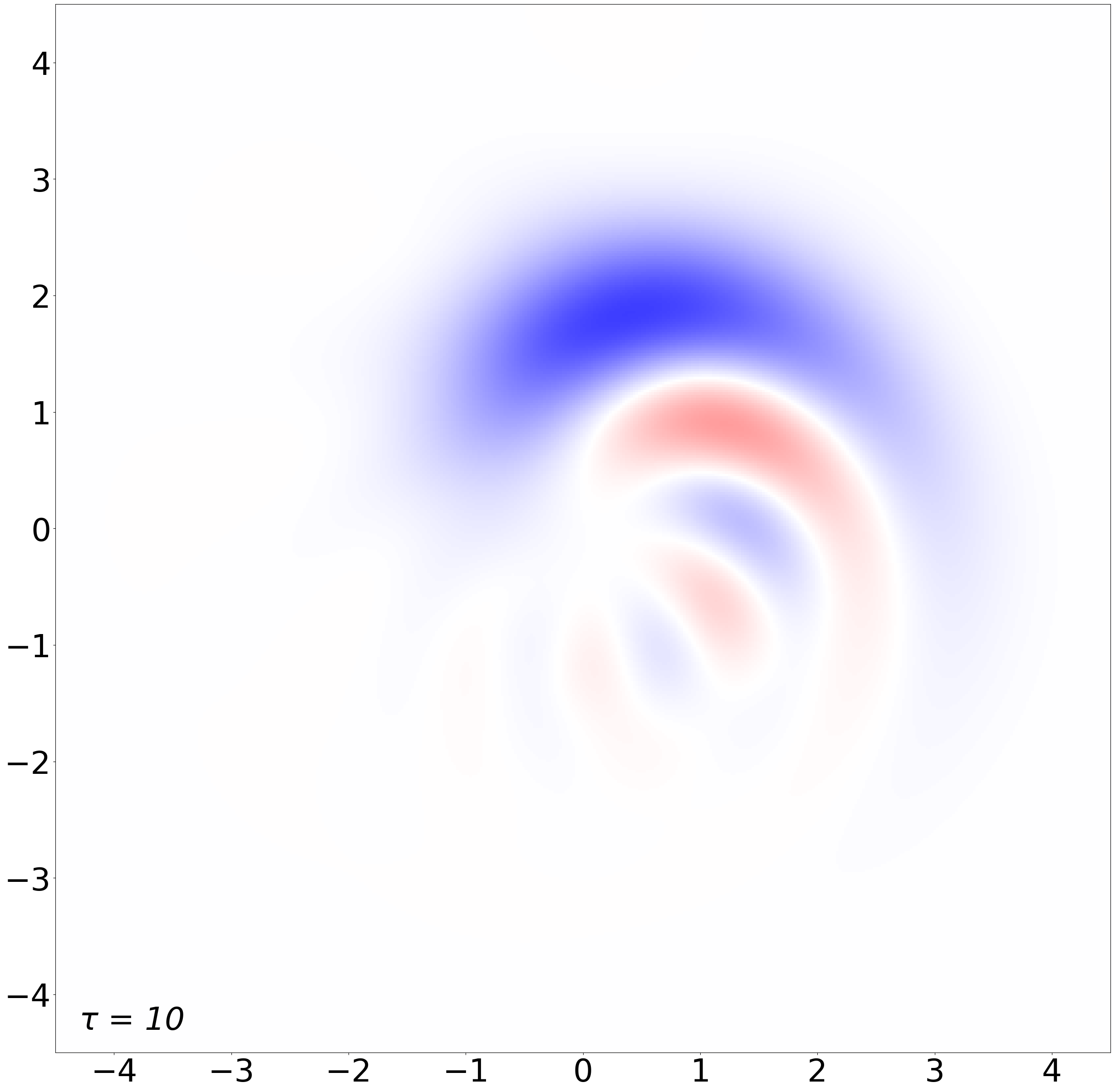}
\caption{$t=10$}
\label{fig: 2(b)}
\end{subfigure}
\begin{subfigure}{0.32\textwidth}
\includegraphics[height=4cm,width=4cm]{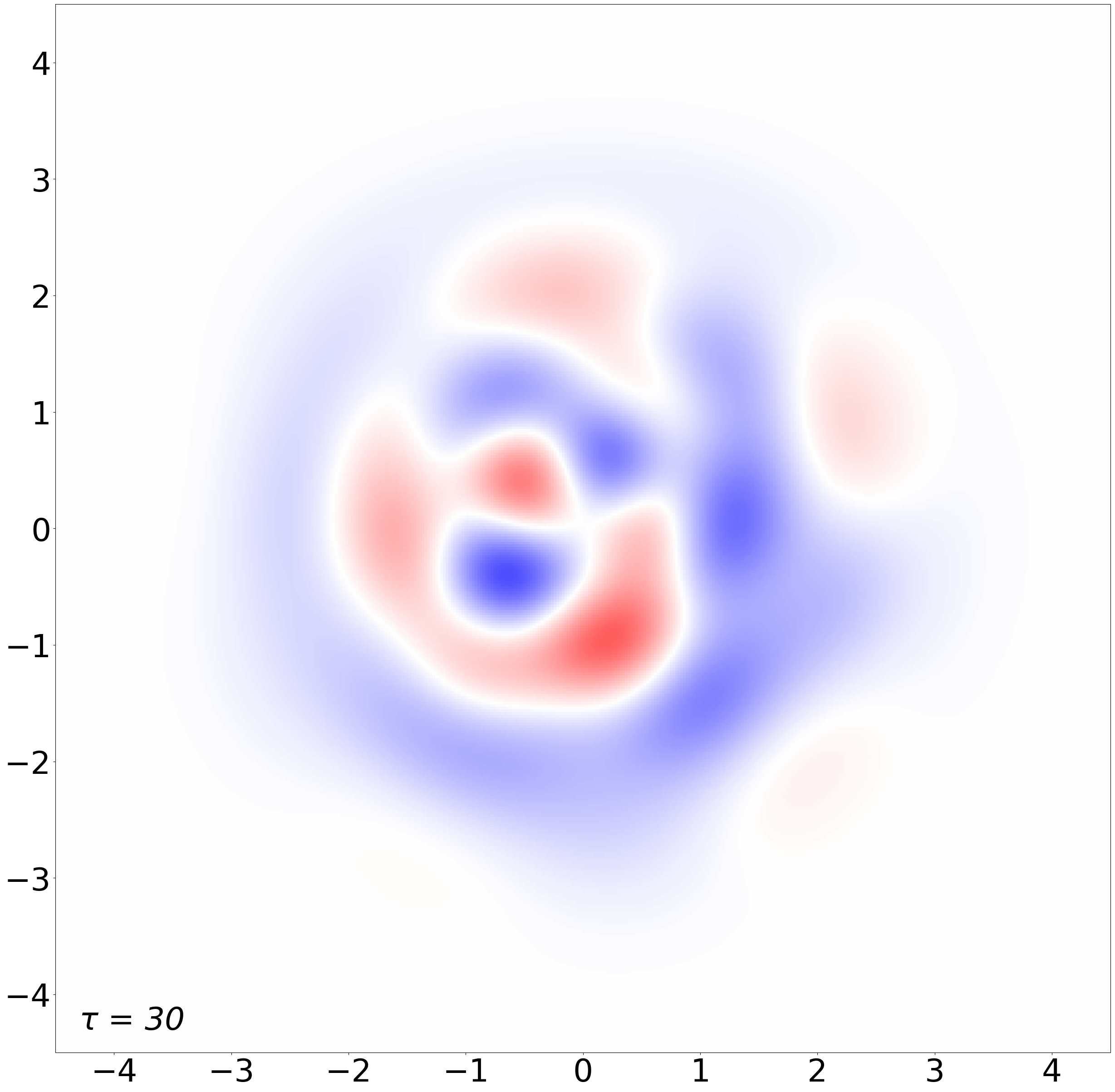}
\caption{$t=30$}
\label{fig: 2(c)}
\end{subfigure}
\begin{subfigure}{0.32\textwidth}
\includegraphics[height=4cm,width=4cm]{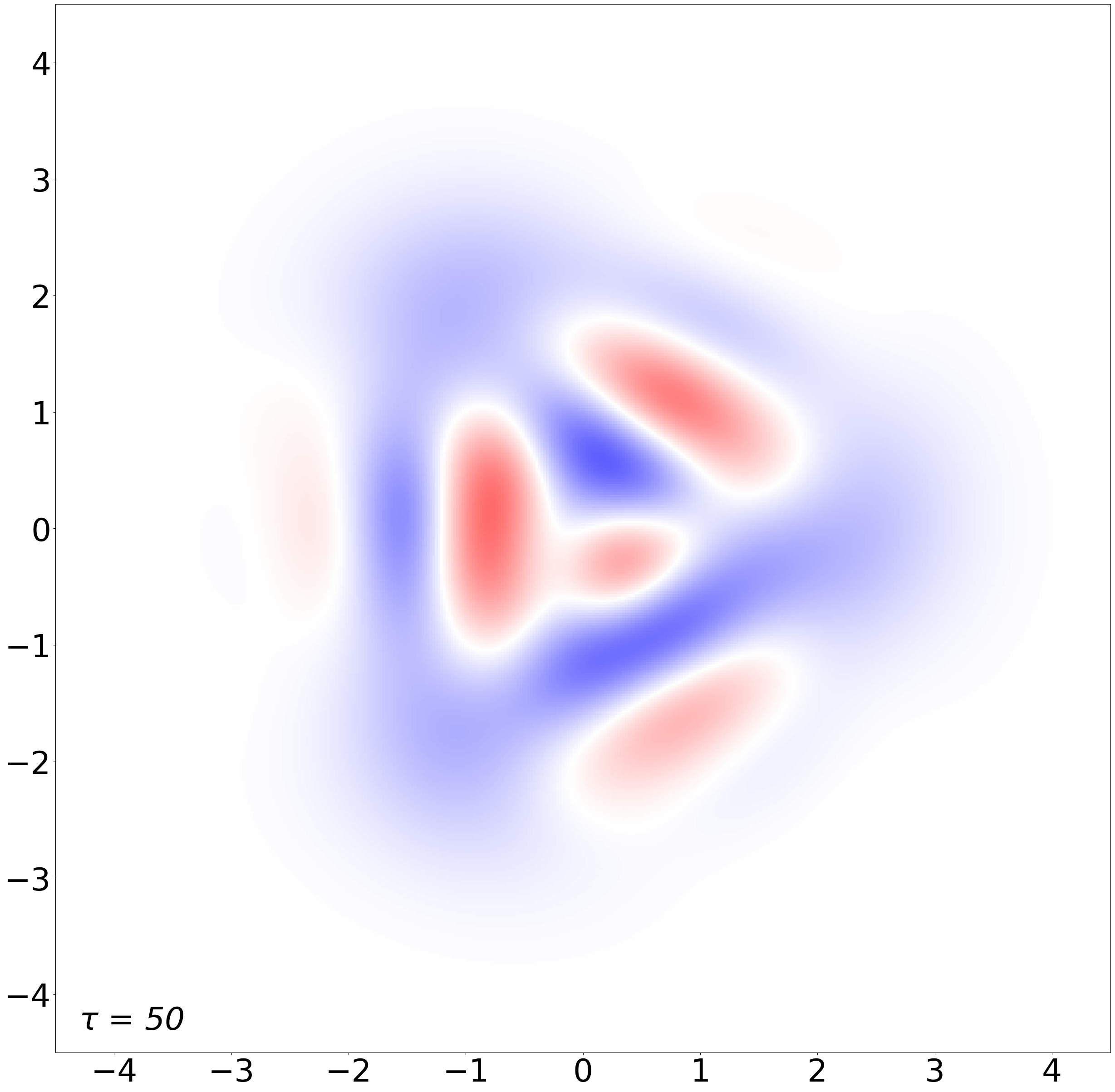}
\caption{$t=50$}
\label{fig: 2(d)}
\end{subfigure}
\begin{subfigure}{0.32\textwidth}
\includegraphics[height=4cm,width=4cm]{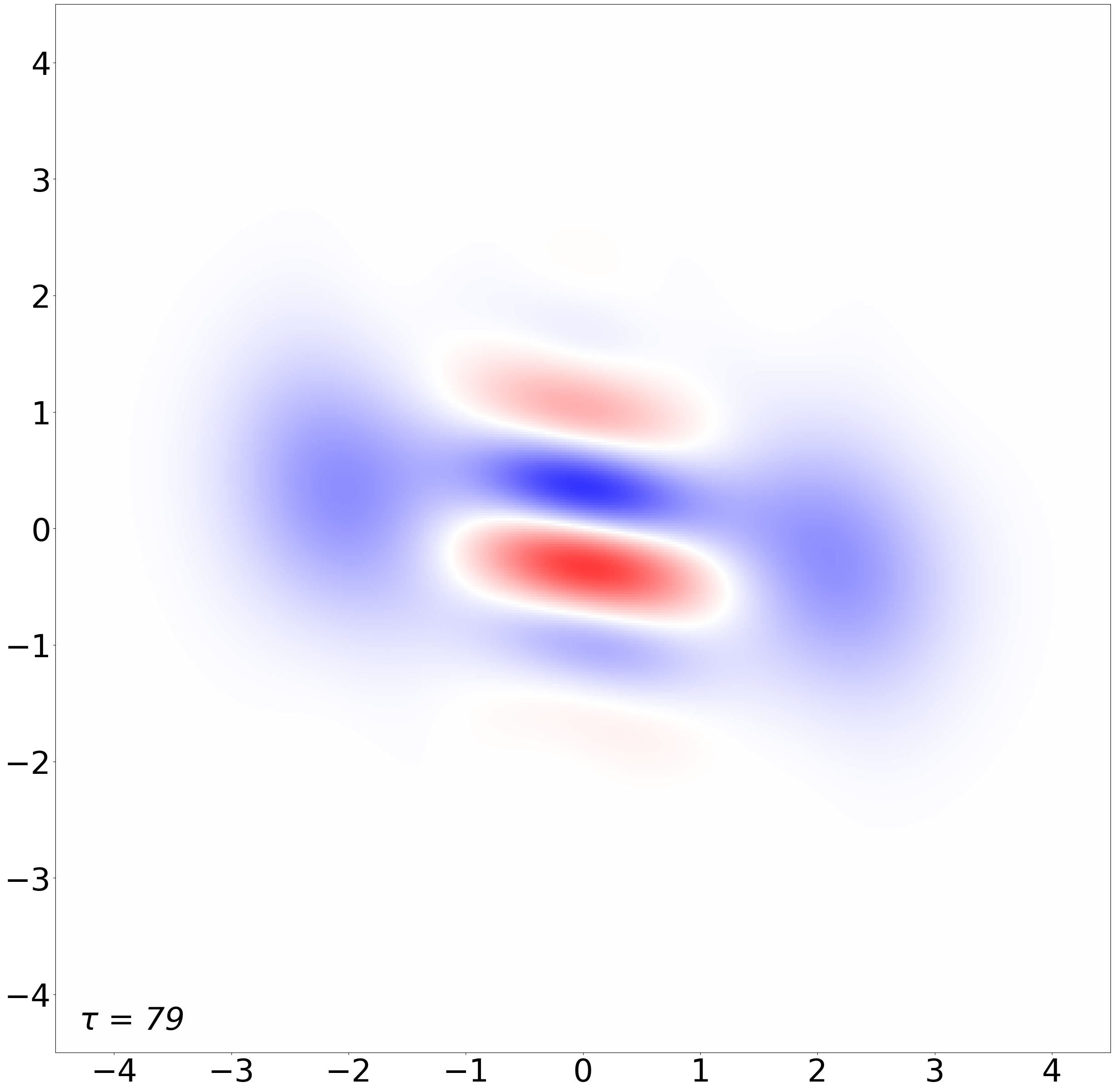}
\caption{$t=79$}
\label{fig: 2(e)}
\end{subfigure}
\begin{subfigure}{0.32\textwidth}
\includegraphics[height=4cm,width=4cm]{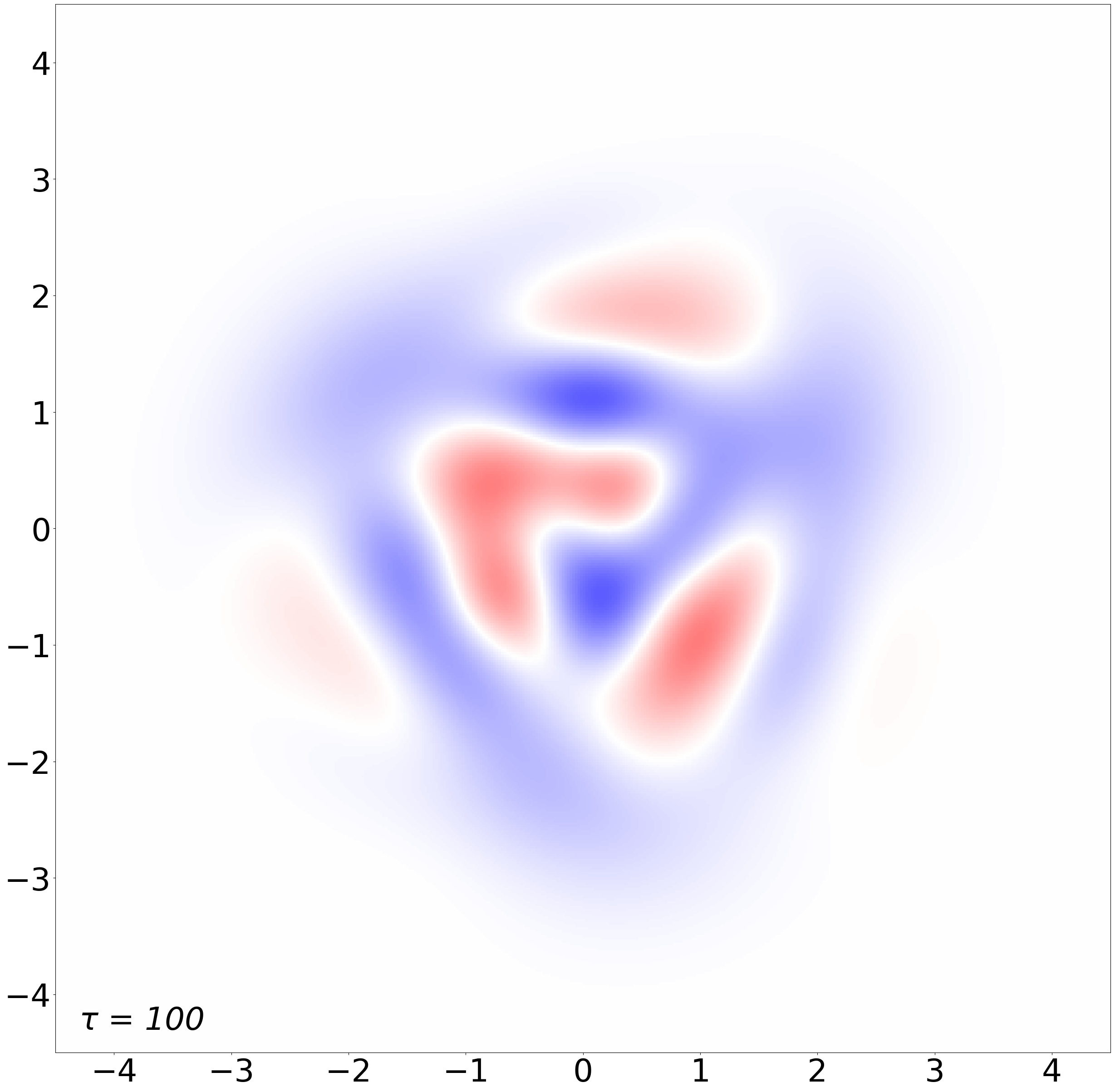}
\caption{$t=100$}
\label{fig: 2(f)}
\end{subfigure}
\begin{subfigure}{0.32\textwidth}
\includegraphics[height=4cm,width=4cm]{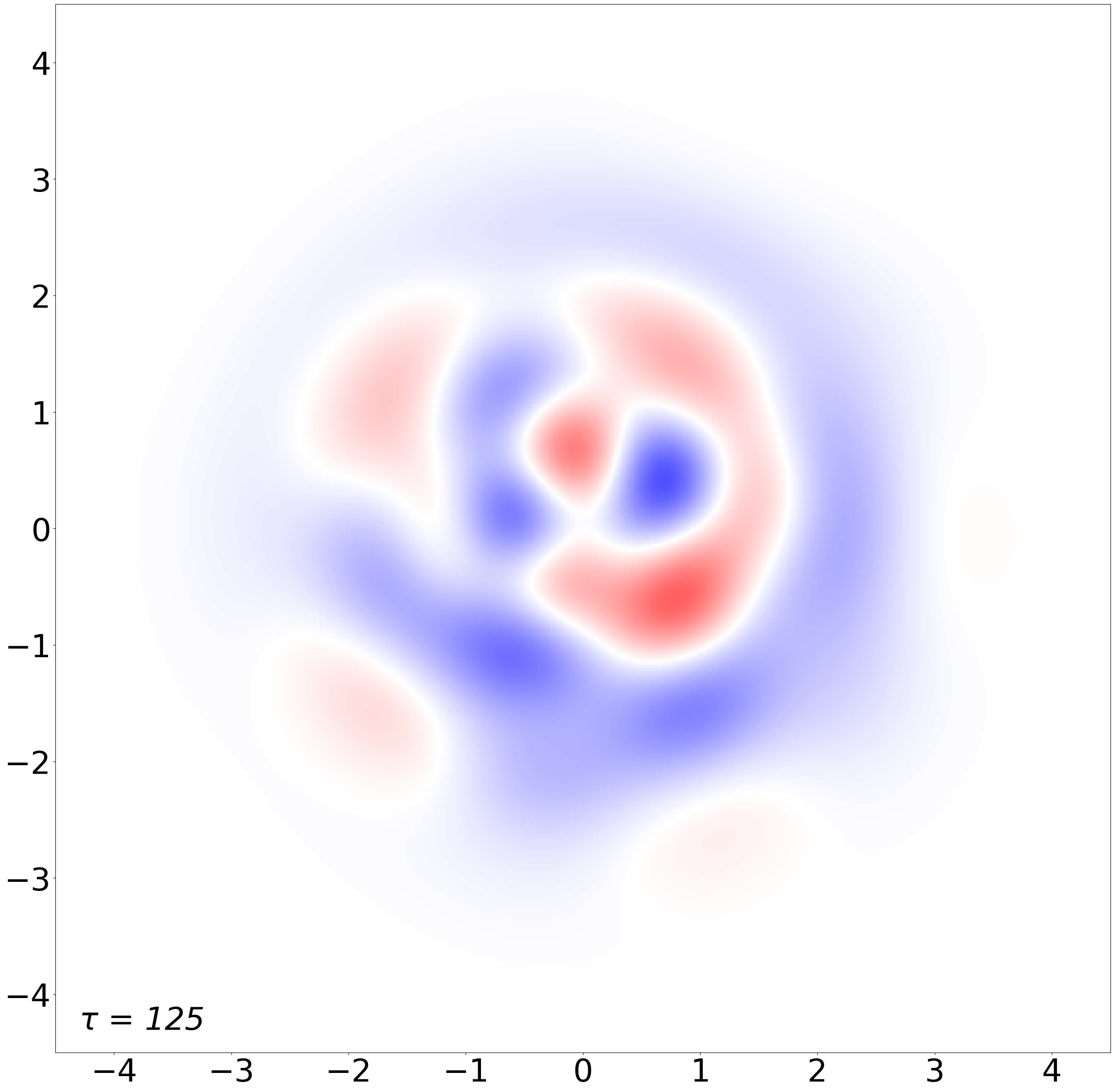}
\caption{$t=125$}
\label{fig: 2(g)}
\end{subfigure}
\begin{subfigure}{0.32\textwidth}
\includegraphics[height=4cm,width=4cm]{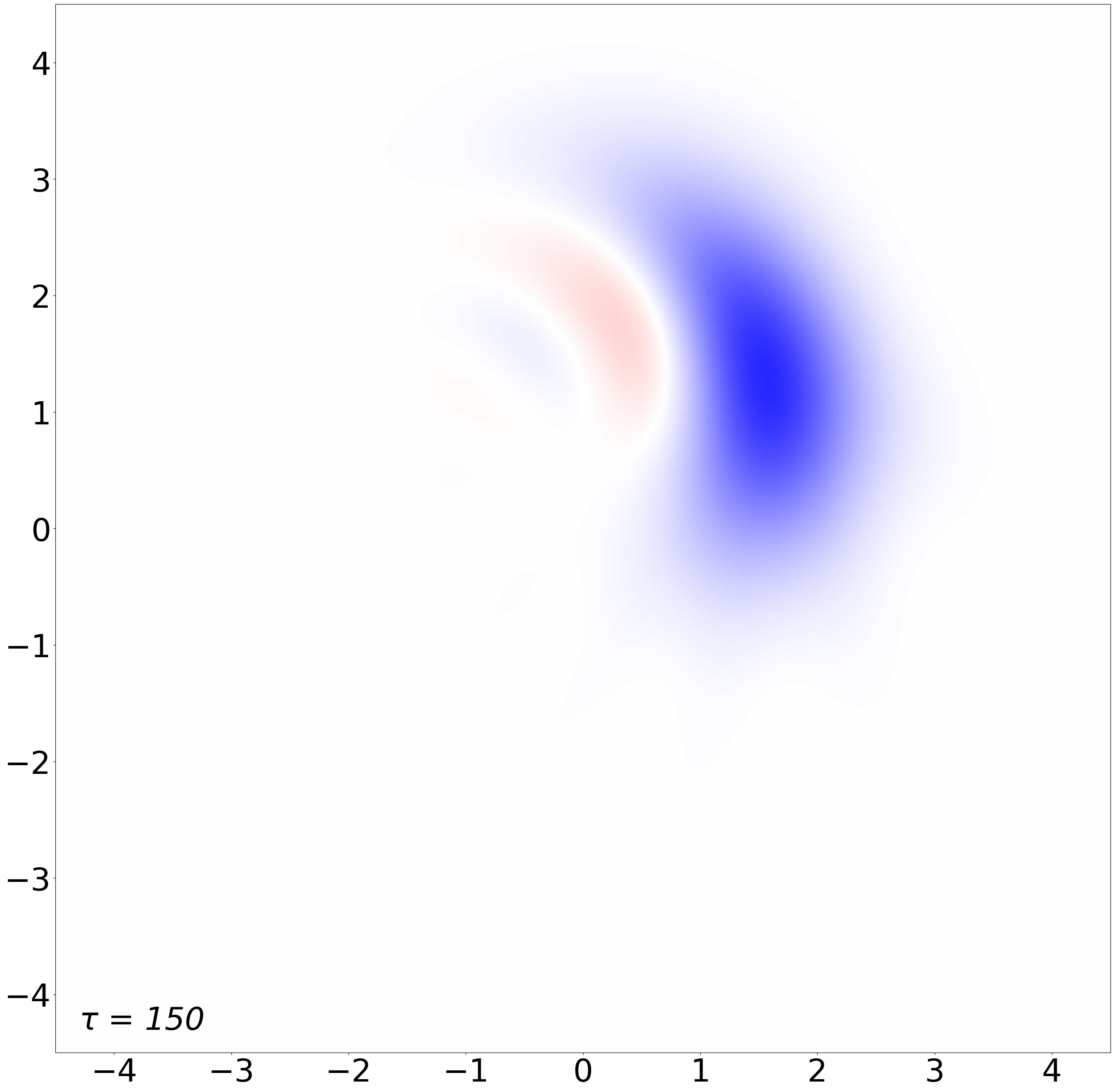}
\caption{$t=150$}
\label{fig: 2(h)}
\end{subfigure}
\begin{subfigure}{0.32\textwidth}
\includegraphics[height=4cm,width=4cm]{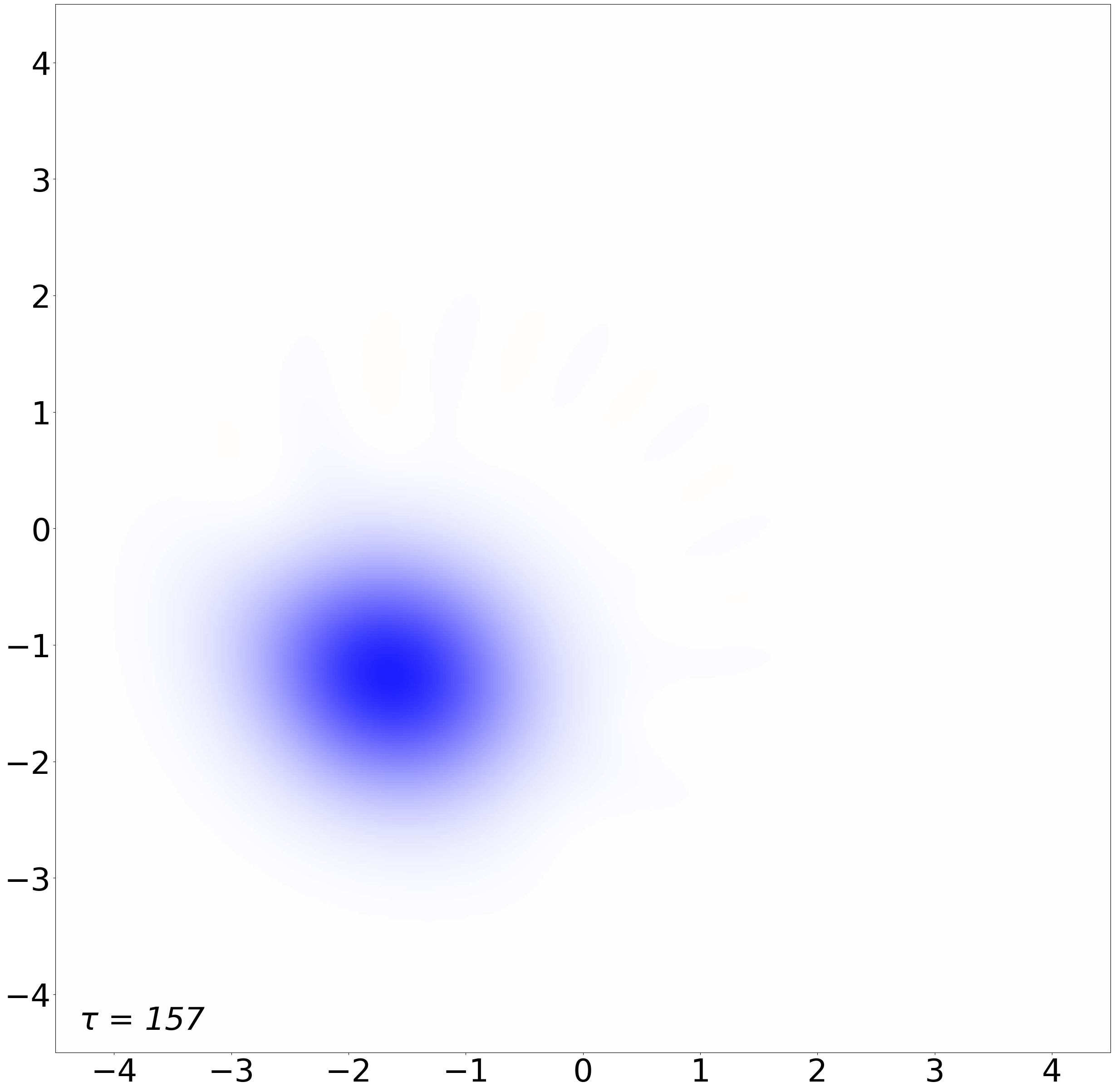}
\caption{$t=157$}
\label{fig: 2(i)}
\end{subfigure}
\begin{subfigure}{0.32\textwidth}
\includegraphics[height=4cm,width=4cm]{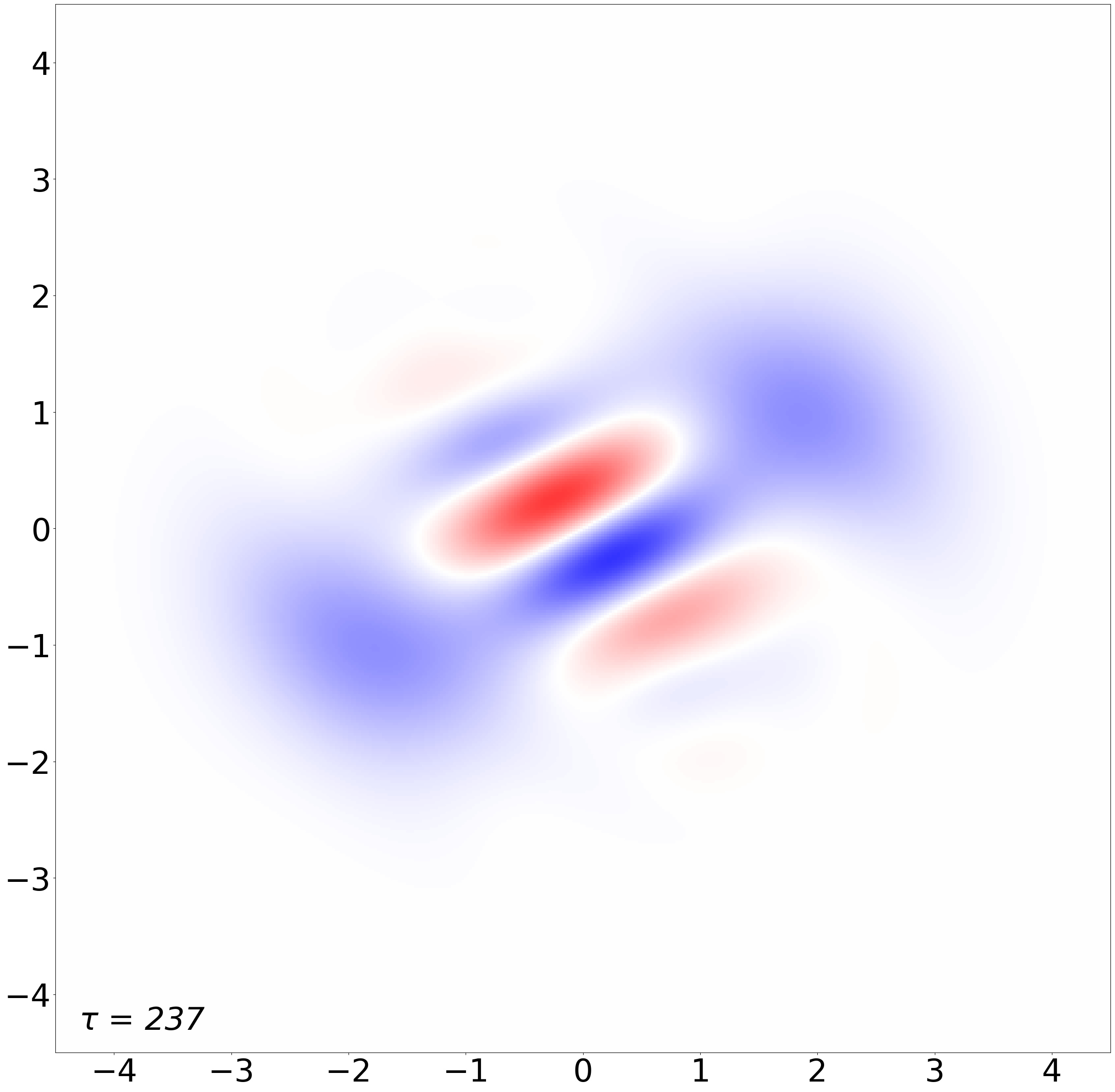}
\caption{$t=237$}
\label{fig: 3(d)}
\end{subfigure}
\begin{subfigure}{0.32\textwidth}
\includegraphics[height=4cm,width=4cm]{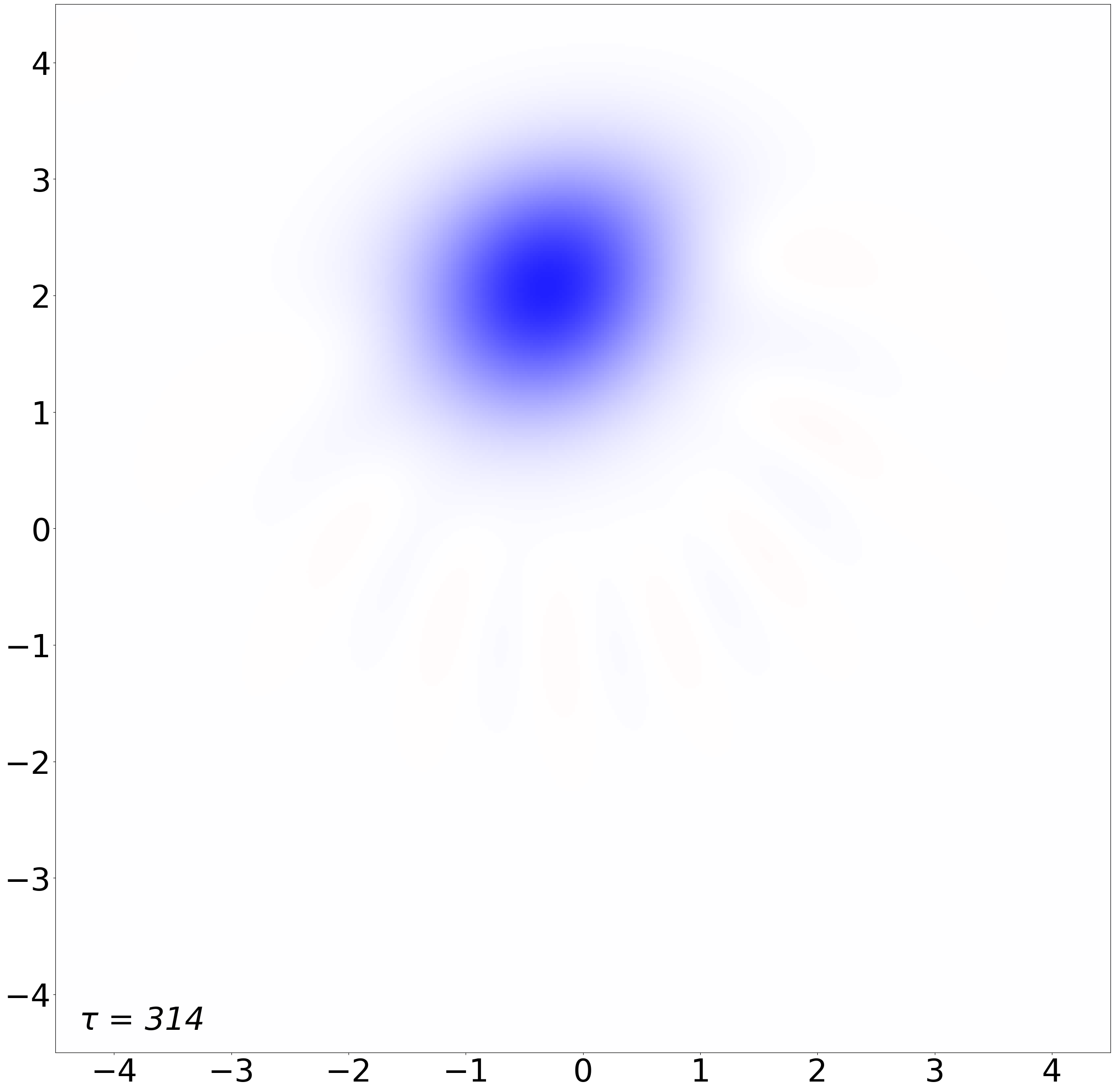}
\caption{$t=314$}
\label{fig: 3(e)}
\end{subfigure}
\begin{subfigure}{0.32\textwidth}
\includegraphics[height=4cm,width=4cm]{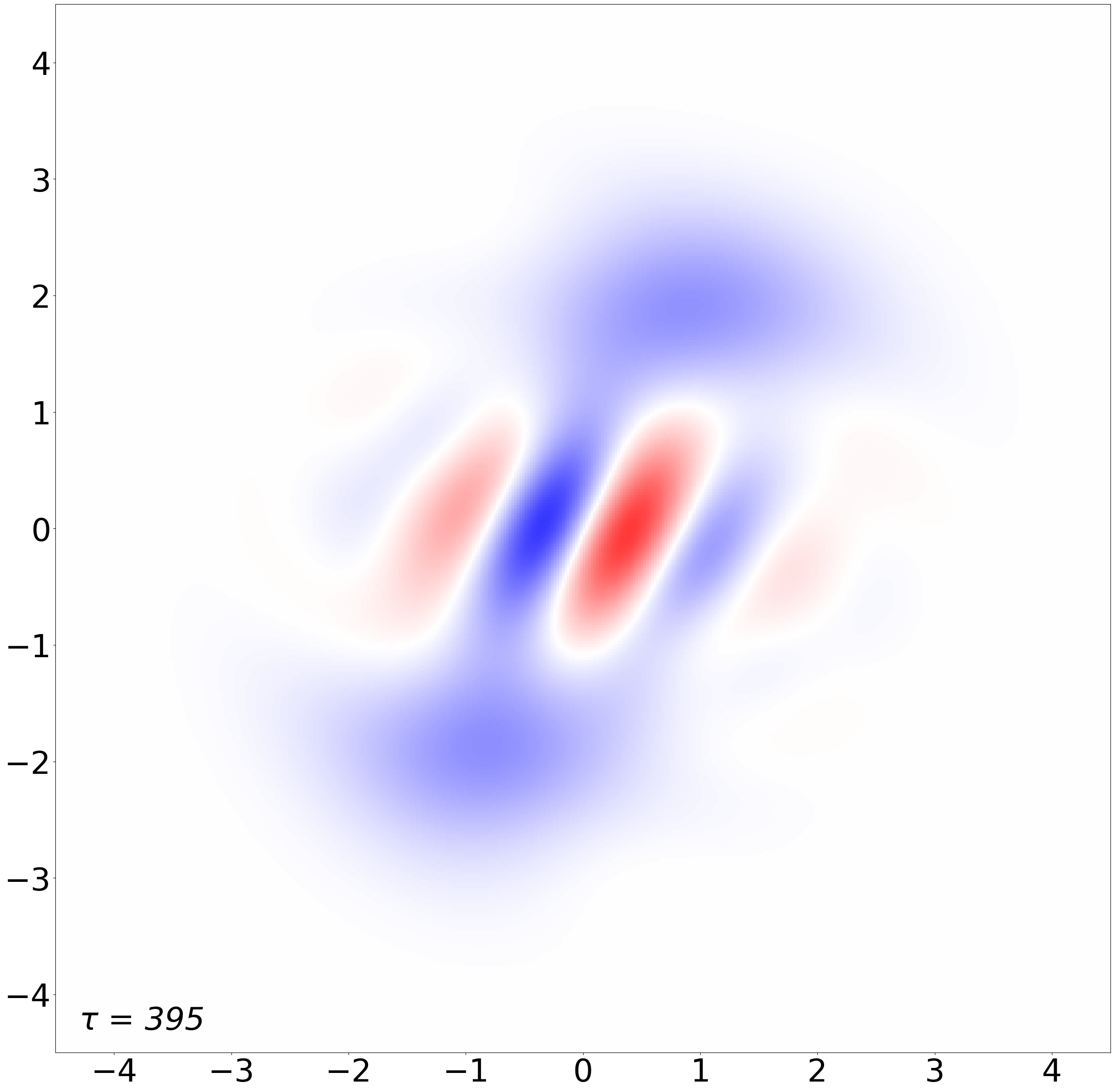}
\caption{$t=395$}
\label{fig: 3(f)}
\end{subfigure}
\begin{subfigure}{0.32\textwidth}
\includegraphics[height=4cm,width=4cm]{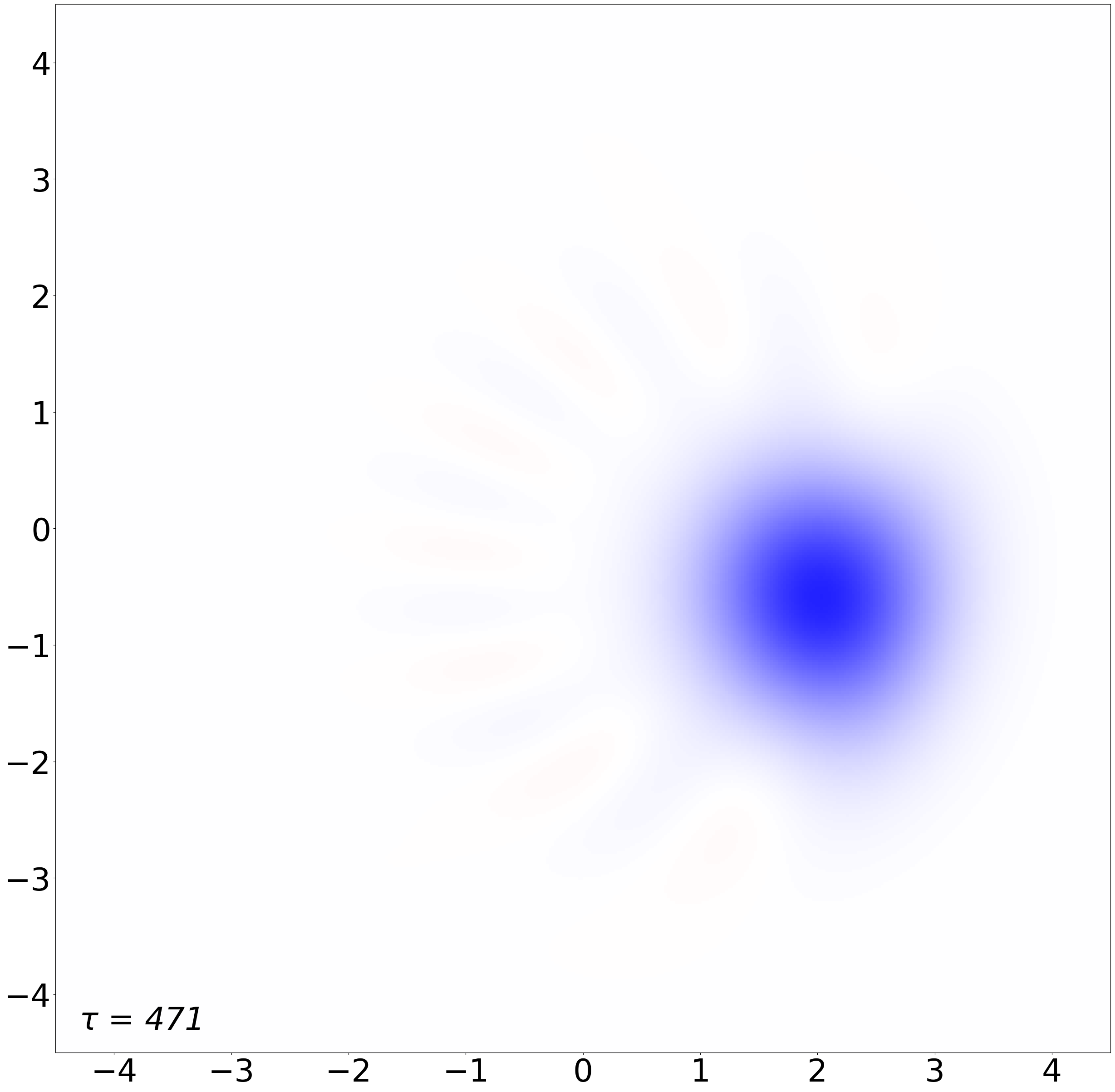}
\caption{$t=471$}
\label{fig: 3(g)}
\end{subfigure}
\begin{subfigure}{0.32\textwidth}
\includegraphics[height=4cm,width=4cm]{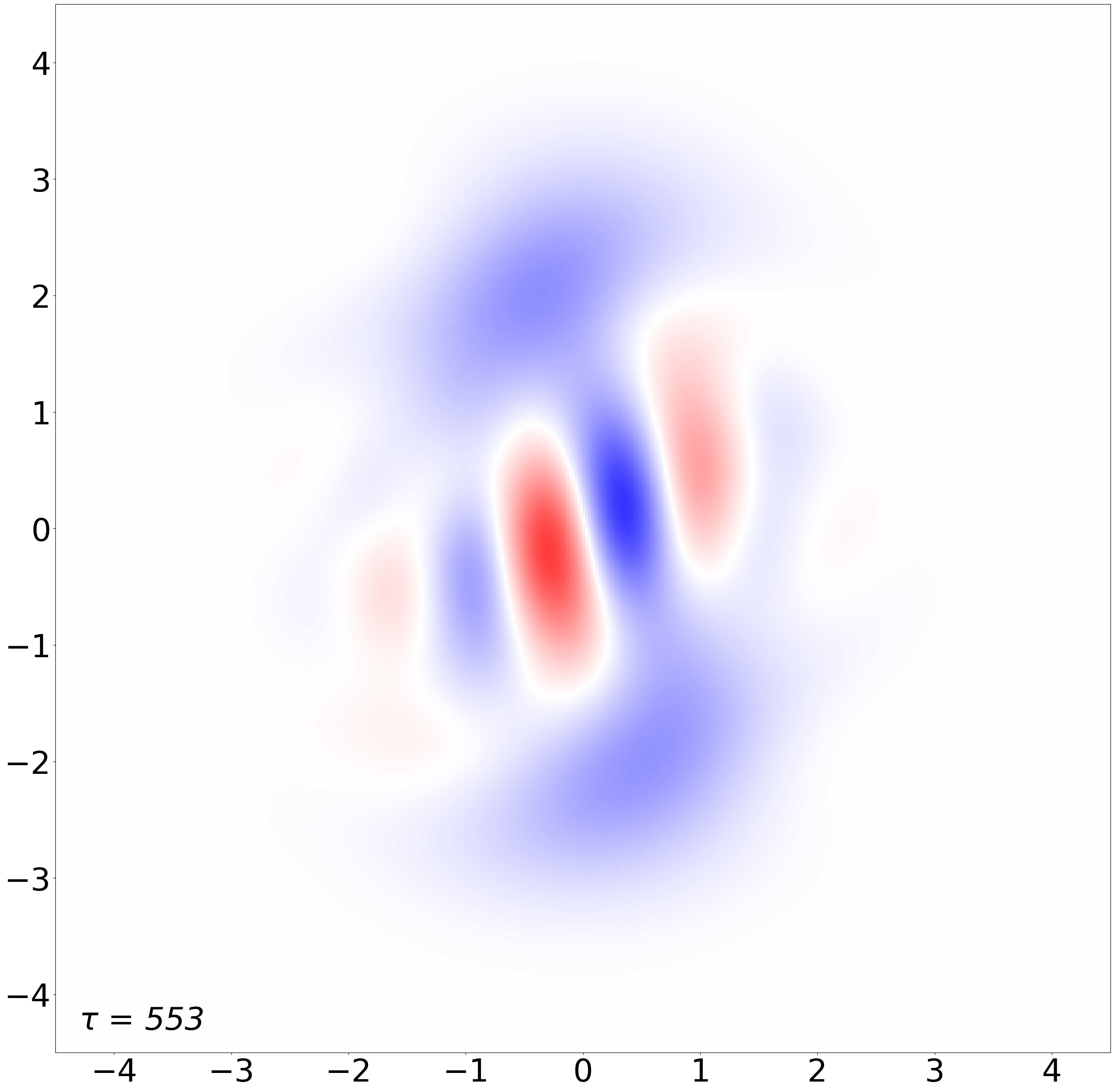}
\caption{$t=553$}
\label{fig: 3(h)}
\end{subfigure}
\begin{subfigure}{0.32\textwidth}
\includegraphics[height=4cm,width=4cm]{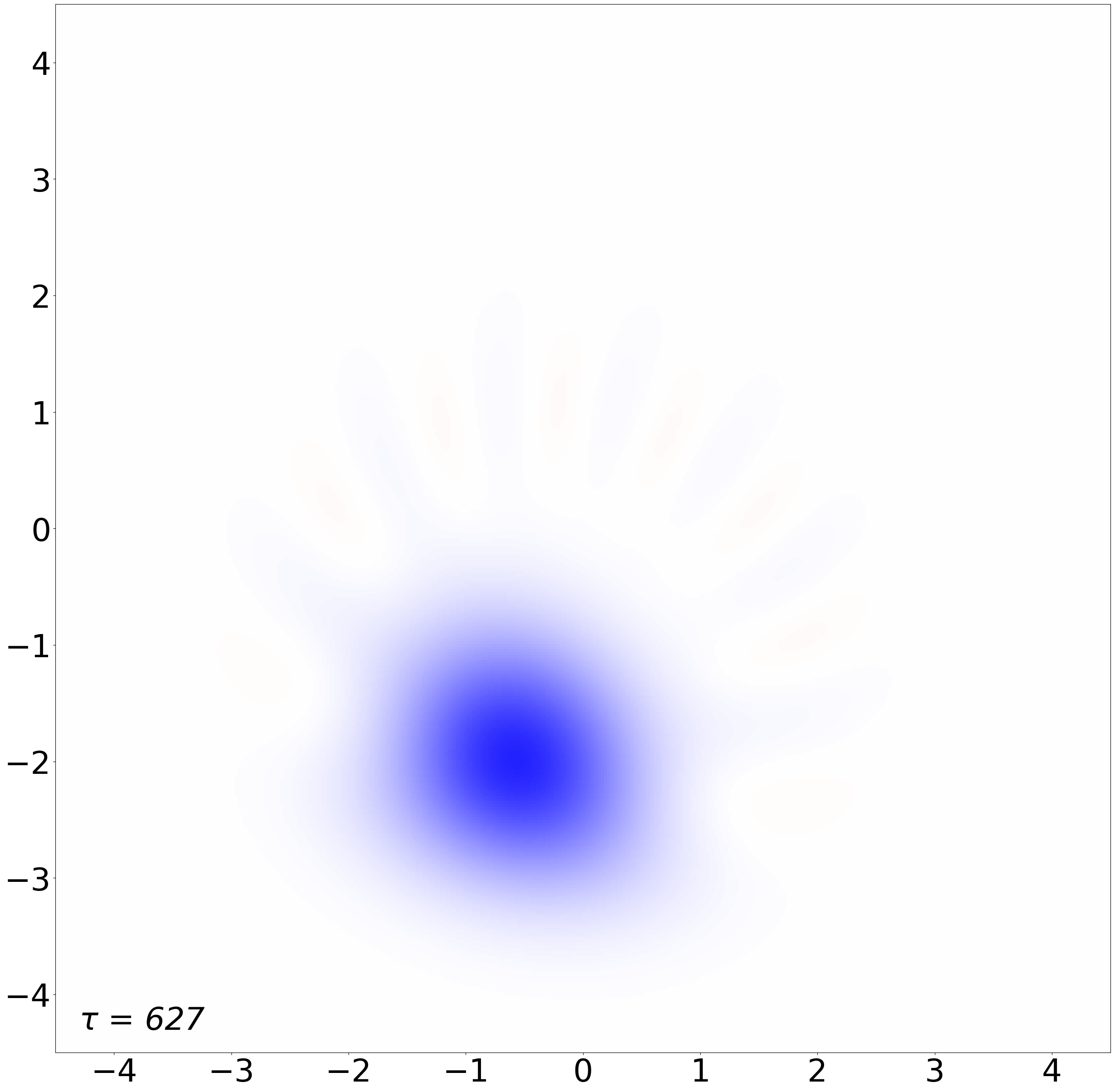}
\caption{$t=627$}
\label{fig: 3(i)}
\end{subfigure}
\caption{Snapshots in term of contour plot of the Wigner
function of the evolving coherent state prepared inside
optomechanical quantum memory.  for N=10. The optical cavity
decay rate $\gamma_{c}$ and the mechanical damping rate
$\gamma_{m} = 0.00001$ a.u.  where the bath temperature
corresponding to each mode is $T=0$.  Non-linearities
present in both the modes take the value $k_{c}$ = $k_{m} =
0.01$~a.u..}
\label{fig:Figure2}
\end{figure*}
\subsection{Master Equation and Phase-space representation
of quantum states}
\label{2(b)}
We develop the theoretical description to model the
decoherence effects on our two mode  open
system.
The environment-system interactions  are assumed
to be Markovian.  In this formalism the time derivative of
the density {operator}  $\rho$ describing the
 quantum state of the non-linear optomechanical
system is given by {the} Master equation
\begin{equation}  \label{eq:8}
\frac{d\rho}{dt} = \mathcal{L}\rho
\end{equation}
where $\mathcal{L}$ is the Lindblad operator 
acting on the density matrix $\rho$.
It is convenient to decompose $\mathcal{L}$ into 
\begin{equation} \label{eq:9}
\mathcal{L} = \mathcal{L}_{c}+\mathcal{L}_{m}+\mathcal{L}_{int}
\end{equation}
where the operators $\mathcal{L}_{c}$ and
$\mathcal{L}_{m}$ act on the optical mode and 
mechanical mode, respectively while the operator
$\mathcal{L}_{int}$  
describes coupling in the optomechanical system.

On the account of $H_c$ in equation~(\ref{eq:2}), the total
Lindblad operator for optical cavity is given as:
\begin{equation} \label{eq:12}
\mathcal{L}_{c}\rho=-\frac{\iota}{\hbar}[\omega_{c} a^{\dagger}
a+k_{c}(a^{\dagger}a)^{2},\rho] + L_{c}\rho,
\end{equation}
where $L_c$ describes the collapse operator as
\begin{equation}  \label{eq:10}
\begin{split} 
L_{c} \rho = \gamma_{c} (n_{c}+1)(a \rho
a^{\dagger}-\frac{1}{2}(a^{\dagger}a \rho+\rho
a^{\dagger}a))\\
+  \gamma_{c} (n_{c})(a^\dagger \rho a -\frac{1}{2}(a \rho
a^{\dagger}+\rho a a^{\dagger})).
\end{split}
\end{equation} 
Here $n_{c}$ is the mean photon occupation number of the
surrounding  thermal reservoir kept at temperature $T$
associated with optical mode and is given by
\begin{equation} \label{eq:11}
n_{c}=\frac{1}{exp(\frac{\hbar w_c}{K_B T})-1}.
\end{equation}
In analogy to equation~(\ref{eq:12}), for the 
mechanical mode $b$ 
associated with mechanical oscillator, 
the total Lindblad operator is given by
\begin{equation} \label{eq:13}
\mathcal{L}_{m}\rho=-\frac{\iota}{\hbar}[\omega_{m} b^{\dagger}
b+k_{m}(b^{\dagger}b)^{2},\rho] + L_{m}\rho,
\end{equation}
where 
\begin{equation} \label{eq:14}
\begin{split} 
L_{m} \rho = \gamma_{m} (n_{m}+1)(b \rho
b^{\dagger}-\frac{1}{2}(b^{\dagger}b \rho+\rho
b^{\dagger}b))\\
+  \gamma_{m} (n_{m})(b^\dagger \rho b -\frac{1}{2}(b \rho
b^{\dagger}+\rho b b^{\dagger}))
\end{split}
\end{equation}
and the mean thermal phonon occupation number of the bath
(at temperature $T$) 
corresponding to mechanical mode is
\begin{equation} \label{eq:15} 
n_{m}=\frac{1}{exp(\frac{ \hbar w_m}{K_B T})-1}.
\end{equation}
Eventually, the coupling Hamiltonian $H_{coupling}$  in
equation~(\ref{eq:2}) couples the number of photons
$\hat{n}=a^{\dagger}a$ of the optical mode to the position
$\hat{x}=b+b^{\dagger}$ of the mechanical mode.  Its
Lindblad operator is denoted as
\begin{equation} 
\label{eq:16}
\mathcal{L}_{int}\rho=
-\frac{\iota}{\hbar}[-g_{0}a^{\dagger}a(b^{\dagger}+b),\rho].
\end{equation}
As the Lindblad Master equation (equation~(\ref{eq:8})) describes
the quantum dynamics of the system state characterized by
density {operator}  $\rho$(t), we use
Wigner function to {describe the quantum state in phase
space}~\cite{PhysRevA.96.013854}.  Wigner
function is a phase-space representation whose integration
with respect to the system position coordinate $x$
gives the marginal probability density in the momentum
coordinate $p$ (and vice versa)
~\cite{PhysRevA.100.012124}.  The Wigner function
representation of the quantum state $\rho$(t) on phase-space
{is} defined
as~\cite{wigner1997quantum,hillery1984distribution,case2008wigner,curtright2013concise,PhysRevA.100.012124}
\begin{equation} \label{17}
\begin{split}
W(x,p,t)=\frac{1}{\pi\hbar}\int_{-\infty}^{+\infty} dy\,
e^{-2ipy/\hbar}\langle x+y|{\rho}(t)|x-y\rangle  \\
=\frac{1}{\pi\hbar}\int_{-\infty}^{+\infty} dp'\, e^{+2ip'
x/\hbar}\langle p+p'|{\rho}(t)|p-p'\rangle,
\end{split}
\end{equation}
where the  normalized wave function in momentum space is
proportional to Fourier transform of the wave function in
position space.

Further, to {locate} the 
time {instant} at which it is most appropriate to
retrieve the stored information, we would like to calculate
the  expectation values of amplitude
$\langle$a(t)$\rangle$ for the stored coherent states
{as function of time}.  The solution of
equation~(\ref{eq:8}) for optomechanical system can be written
as
\begin{equation} \label{eq:18}
\rho(t)=exp[\mathcal{L}(t)]\rho(t_{0}).
\end{equation}
From the above equation one can compute the time evolved
density operator $\rho$(t) which is
further used to calculate the 
expectation
value of the amplitude via the relation given by
\begin{equation}\label{eq:19}
\langle a(t) \rangle=\langle a |\rho(t)| a \rangle
\end{equation}
We calculate and display the snapshots of Wigner 
function at different times and plots of 
the expectation of the amplitude as two indicators 
in our analysis.
%%%%%%%%%%%%%%%%%%%%%%%%%%%%%%%%%%%%%%%%%%%%%%%%
\begin{figure} 
\includegraphics[height=5cm,width=9cm]{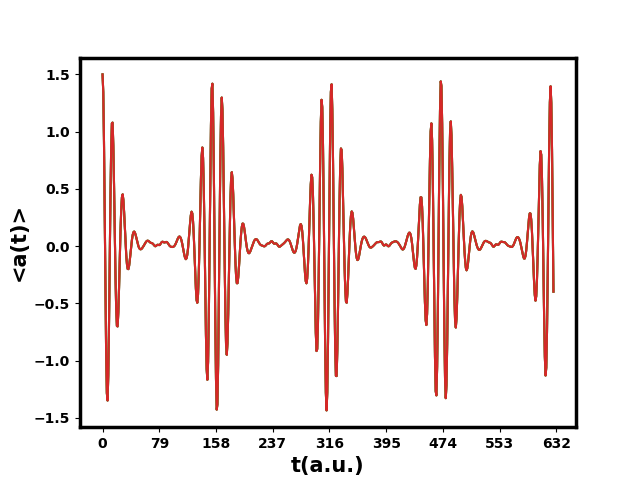}
\caption{The expectation value $\langle a(t) \rangle$
plotted as a function of time for  coherent quantum state in
the presence of dissipation $\gamma_c=\gamma_m=0.00001$~a.u.
and non linearities $k_c=k_m= 0.01$~a.u. . The bath
temperature corresponding to each mode is $T=0$.}
\label{fig:Figure 4}
\end{figure}

\subsection{Details of numerical simulations} 
\label{2(c)}
All numerical simulations {were} carried out using
QuTiP~\cite{JOHANSSON20121760} which is an open source
software package written in Python.  The numerical results
are obtained by first solving the Lindblad master equation
(equation~(\ref{eq:8})) to get the time evolved coherent state
density matrix.  The density evolution is further extended
to Wigner function evolution after adopting the complete
parameters of the corresponding phase space.
{Further,}  the
expectation value of {the amplitude, representing
quantum information} is obtained by
solving equation~(\ref{eq:19}).  The evolution of the stored
information has been observed until 2$T_{rev}$ where
$T_{rev}$ is given by equation~(\ref{eq:7}).

%Optomechanical system is initially studied in $N=10$ Hilbert space. 
To choose the parameters for our system that show
experimental consistency, we follow the work done by
~\cite{Chakraborty:17}. They numerically illustrated the
effect of cross-Kerr coupling on the steady-state behavior
and stability condition of the optomechanical system by
considering the experimentally accessible frequencies of
optical and mechanical modes as $\omega_{c}= 2\pi\times 370$
THz and $\omega_{m}=2\pi\times 10$ MHz respectively.  To
embody the strong coupling regime between the two modes, the
value of $g_0 $ is taken as 13.47 THz.  
All the physical quantities are evaluated in atomic units
(a.u.) by substituting $\hbar$ and $K_{B}$ equal to 1. Accordingly, our
corresponding parameters in~a.u. are  $\omega_{c}= 2\pi
\times0.056233$, $\omega_{m}= 2\pi \times
0.151983\times10^{-8}$  and $g_{0}=0.20472\times10^{-2}$.
%%%%%%%%%%%%%%%%%%%%%%%%%%%%%%%%%%
\begin{figure*} 
\begin{subfigure}{0.45\textwidth}
\includegraphics[height=5cm,width=8.5cm]{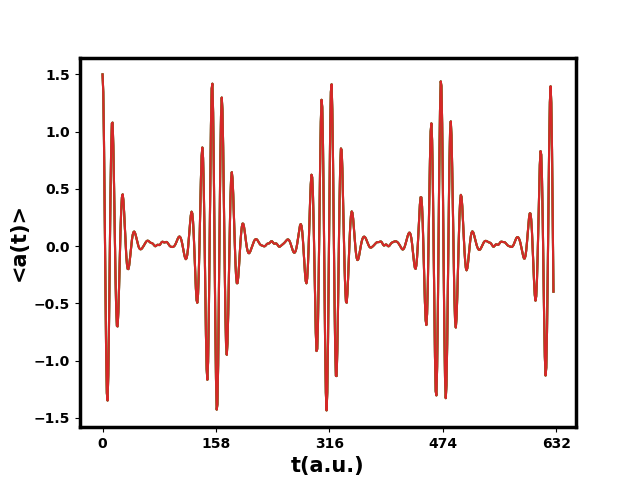}
\caption{Dissipation$= 0.00001$, Non-linearity$= 0.01$}
\label{fig:5(a)}
\end{subfigure}
\begin{subfigure}{0.45\textwidth}
\includegraphics[height=5cm,width=8.5cm]{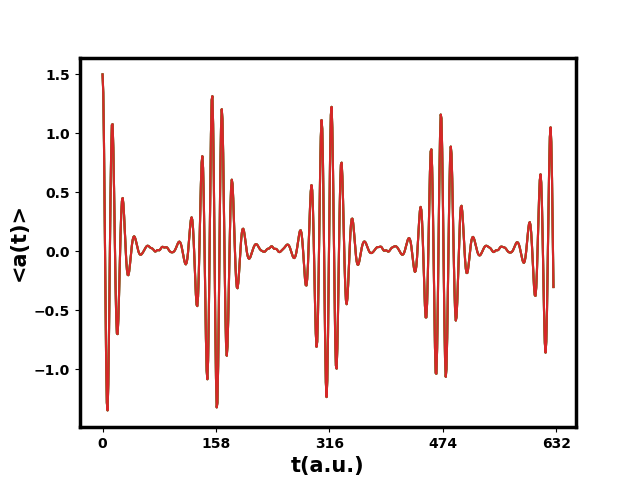}
\caption{Dissipation$= 0.0001$, Non-linearity$= 0.01$}
\label{fig:5(b)}
\end{subfigure}
\begin{subfigure}{0.45\textwidth}
\includegraphics[height=5cm,width=8.5cm]{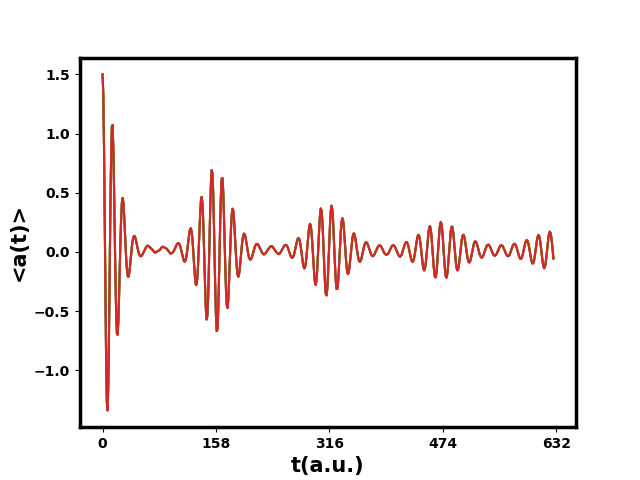}
\caption{Dissipation$= 0.001$, Non-linearity= $0.01$}
\label{fig:5(c)}
\end{subfigure}
\begin{subfigure}{0.45\textwidth}
\includegraphics[height=5cm,width=8.5cm]{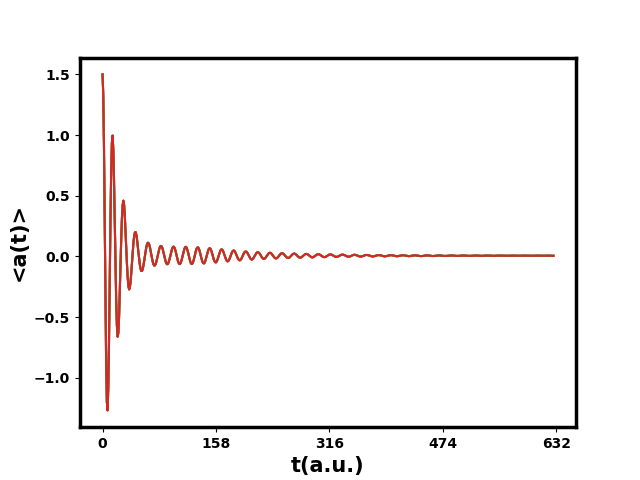}
\caption{Dissipation= $0.01$, Non-linearity= $0.01$}
\label{fig:5(d)}
\end{subfigure}
\caption{Collapse and revival of a coherent quantum
state inside an optomechanical oscillator with non
linearities $k_c=k_m= 0.01$~a.u.  for four different values
of dissipation. (a)$\gamma_c = \gamma_m =0.00001 $~a.u.,
(b)$\gamma_c = \gamma_m =0.0001 $~a.u., (c)$\gamma_c =
\gamma_m =0.001 $~a.u.  and (d)$\gamma_c = \gamma_m =0.01 $
a.u..}
\label{fig:Figure 5}
\end{figure*}

\section{Results and Discussion}
 \label{3}
%\subsection{Wigner Representation}
\subsection{Visualization of the time evolution through
Wigner representation}
%{\color{red} {\bf Note:} I hve changed to font of all 
%number of math mode, please check if anywhere it is still
%need. I have replaced figure with Figure where ever we are
%referring to figures, again check if any instance is left.
%However, for subfigure references they should appear as
%in Figure 2(a) and not as Figure 2a. This needs to be 
%corrected by understanding the caption package.}
%%%%%%%%%%%%%
We plot the Wigner function corresponding to the quantum
state at different times in order to see the transformations
caused in the state due to nonlinearity and
system-environment coupling. 
We consider an initial bit of quantum information embedded
in a coherent state of amplitude  $\alpha=1.5$. The
effective non-linearity of both the optical and mechanical
modes is taken as $0.01$~a.u. The bath is maintained at 0K
such that $n_{c}=n_{m}=0$. The overall dissipation resulting
from this system-bath interaction is selected to be
$\gamma_{c} = \gamma_{m} = 0.00001$~a.u. It is observed that
during {the time} evolution, coherent 
state undergoes series of alterations as is obvious
in the contour plots of the Wigner function displayed at
different times in Figure~\ref{fig:Figure2}. 
In these plots, (Figure~\ref{fig: 2(a)}-~\ref{fig: 3(i)})
 regions color coded blue depict positive Wigner function
which correspond to conventional probability density while
red regions depict negative Wigner function which correspond
to {a} non-classical states.  
%A unit area square corresponding to Planck’s constant $\hbar$ is indicated at the bottom right of each figure to give the scale in each snapshot of Wigner
representation~\cite{PhysRevA.100.012124}.
%%%%%%%%%%%%%%%%%%%%%%%%%%%%%%

The Wigner plot for the initial coherent state at $t=0$~a.u.,
an elliptical blue region   
is shown in Figure~\ref{fig: 2(a)}. 
%Elliptical blue regions depict
%no effects from the environment thereby preserving its
%information.  
As time advances {the} 
coherent state looses its classicality
and the non-classical red regions appear. The resulting
state is marked by  a tail of interference fringes as
portrayed by snapshot at $t=10$~a.u. (Figure~\ref{fig: 2(b)}).
Further evolution is marked by the appearance of highly
non-classical  multi-component Schr\"{o}dinger cat states,
which are also called kitten states as shown in
Figures~\ref{fig: 2(c)} and \ref{fig: 2(d)}.  Two component
Schr\"{o}dinger cat states are also observed at 
$t=\frac{T_{rev}}{4}=79$~a.u. as depicted in  Figure~\ref{fig: 2(e)}.
During the evolution, appearance of non-classicality in
evolving coherent state is due to natural dispersion.
However, the presence of non-linearities compensate the
dispersion and make the coherent state to 
revive.
Eventually,  initial classical coherent state reappears at t=
$\frac{T_{rev}}{2} =157 $a.u. given in Figure~\ref{fig: 2(i)}
with nonclassicality disappearing!
As we further observe the time evolution of Wigner
function we notice that the coherence revival states
(coherent states) arise at multiples of $\frac{T_{rev}}{2}$
time  as illustrated by Figures~\ref{fig: 2(i)}, ~\ref{fig:
3(e)}, ~\ref{fig: 3(g)} and ~\ref{fig: 3(i)}. 
The analysis
above shows that  the coherences
underlying the coherent state, collapse and revive
periodically which plays an important role in deciding when
to retrieve the information. 
%%%%%%%%%%%%%%%%%%%%%%%%%%%%%%%%%%%%%%%%%%%
\subsection{Evolution of the expectation value of the
amplitude}
In order  to elucidate the loss of amplitude at
coherence collapse time (cat states) and regain of the
amplitude at coherence revival time (coherent states),
we calculate  the expectation value of
coherent state amplitude using the Master equation.  
The time evolution of $\langle a(t) \rangle$
shown in Figure~\ref{fig:Figure 4} endorses the loss in
amplitude odd at multiples of $\frac{T_{rev}}{4}$ time during
collapsed state and regain in amplitude at multiples of
$\frac{T_{rev}}{2}$ during revived state. 

Further to study the extent of degradation of  quantum
information, we extend our results to find the
evolution of the expectation
value of the amplitude under various circumstances.
In particular we study
 (1) the effect of environment, 
 (2) the effect of non-linearities in system,
(3) the effect of bath temperature and  
(4) the effect of initial amplitude of coherent state. 
After this study we
will be able to draw out the best suited combinations for
information retrieval. 
\begin{figure*} 
\begin{subfigure}{0.45\textwidth}
\includegraphics[height=5cm,width=8.5cm]{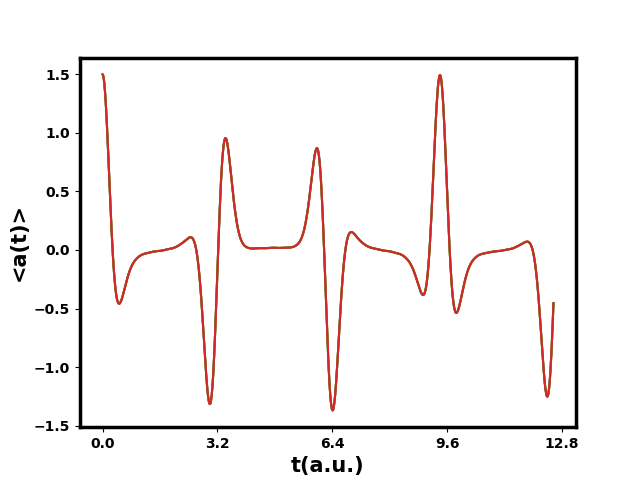}
\caption{Non-linearity$= 0.5$}
\label{fig:6(a)}
\end{subfigure}
\begin{subfigure}{0.45\textwidth}
\includegraphics[height=5cm,width=8.5cm]{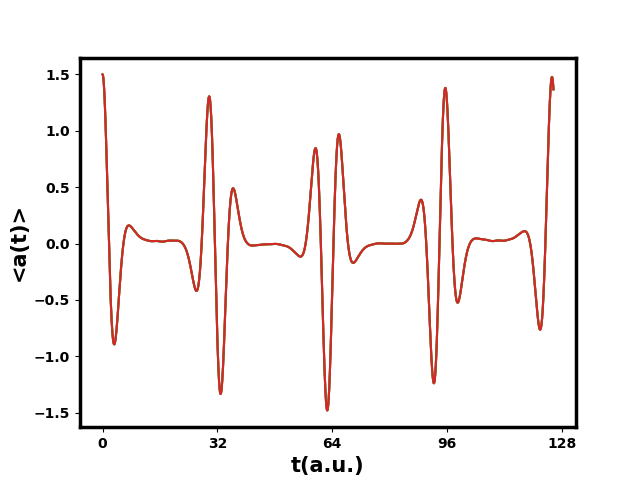}
\caption{Non-linearity$= 0.05$}
\label{fig:6(b)}
\end{subfigure}
\begin{subfigure}{0.45\textwidth}
\includegraphics[height=5cm,width=8.5cm]{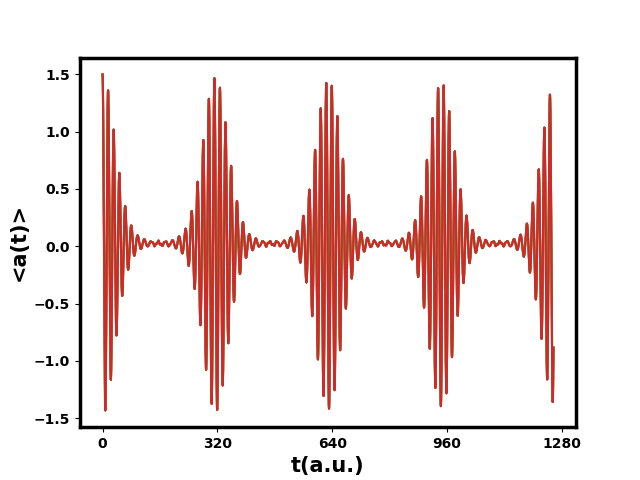}
\caption{Non-linearity$= 0.005$}
\label{fig:6(c)}
\end{subfigure}
\begin{subfigure}{0.45\textwidth}
\includegraphics[height=5cm,width=8.5cm]{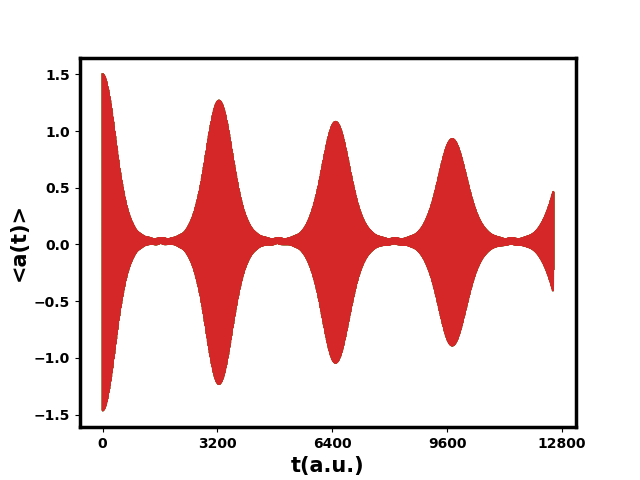}
\caption{Non-linearity$= 0.0005$ }
\label{fig:6(d)}
\end{subfigure}
\caption{Effect of  strong and weak non-linearities in
the optical and mechanical mode represented by $k_c$ and
$k_m$,  respectively,  on the revival behavior for a
coherent quantum state. The dissipations inside the
optomechanical oscillator are $\gamma_c=\gamma_m= 0.00001
a.u.$. The values of non-linearity are (a) $k_c=k_m = 0.5$
a.u., (b) $k_c=k_m = 0.05$~a.u., (c) $k_c=k_m = 0.005$~a.u.
and (d)$k_c=k_m = 0.0005$~a.u.}
\label{fig:Figure 6}
\end{figure*}

%%%%%%%%%%%%%
\subsubsection{ The effect of environment}
%%%%%%%%%%%%%%
We begin with a coherent quantum state and study the
effect of environment on expectation value of its amplitude
as a function of time in the presence of non-linearity.  
We analyze the system in the parameter region
in which the strength of non-linearity
in  the optical mode ($k_c$) as well as
mechanical mode ($k_m$) is $0.01$~a.u. 
The results are displayed in Figure~\ref{fig:Figure 5}
where we have plotted
$\langle a(t) \rangle$ as a function of time and the effect of
environment is studied
for different values of dissipation dignified in terms of
$\gamma_c$ and $\gamma_m$ values which are chosen to be
$0.00001$~a.u., $0.0001$~a.u., $0.001$~a.u. and $0.01$~a.u.  The
dissipation caused by the environment is evident through the
damping and ultimate decay of the expectation value of
amplitude $\langle a(t) \rangle$ as time evolves.  As we scan
through different values of $\gamma$s we find there is a
threshold value of dissipation upto which revivals survive
as illustrated below.

As shown in Figure~\ref{fig:5(a)},  for finite but
small $\gamma_c$ and $\gamma_m$  values (both equal to
$0.00001$~a.u.),   revivals are clearly noticeable although
due to dissipation a reduction in revived amplitude does
take place.  Hence,  although  the revivals are never
complete in the presence of dissipation,  their signatures
are present which can be experimentally identified.  As the
value of $\gamma_c$ and $\gamma_m$ increases the revivals
become weaker and weaker as seen in Figures~\ref{fig:5(b)}
and ~\ref{fig:5(c)}.  From Figure~\ref{fig:5(d)}, one can
conclude that for a certain value of $\gamma_c$ and
$\gamma_m$  equal to $0.01$~a.u., the revivals
disappear altogether.  Hence beyond this threshold value we
do not expect to see revivals in our quantum state.
\begin{figure*}[ht] 
%\begin{subfigure}{0.45\textwidth}
%\includegraphics[height=5cm,width=8.5cm]{tempzero.png}
%\caption{$T = 0$ K}
%\label{fig:7(a)}
%\end{subfigure}
\begin{subfigure}{0.45\textwidth}
\includegraphics[height=5cm,width=8.5cm]{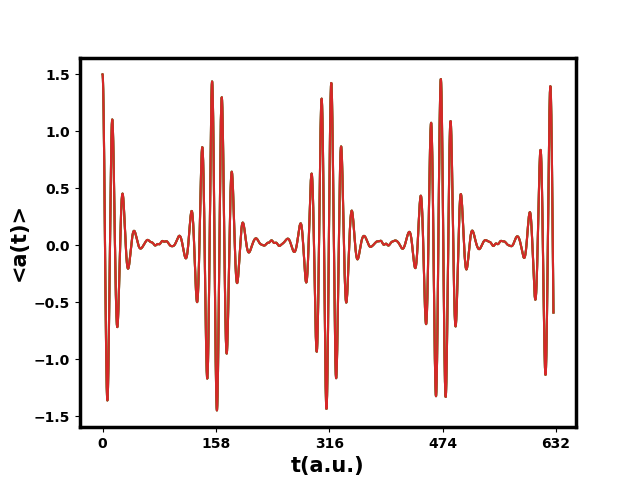}
\caption{$T= 30 \mu$K}
\label{fig:7(b)}
\end{subfigure}
\begin{subfigure}{0.45\textwidth}
\includegraphics[height=5cm,width=8.5cm]{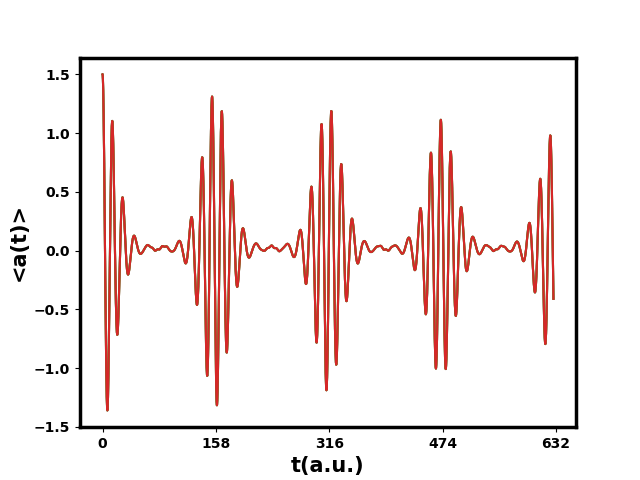}
\caption{$T= 30$ mK }
\label{fig:7(c)}
\end{subfigure}
\begin{subfigure}{0.45\textwidth}
\includegraphics[height=5cm,width=8.5cm]{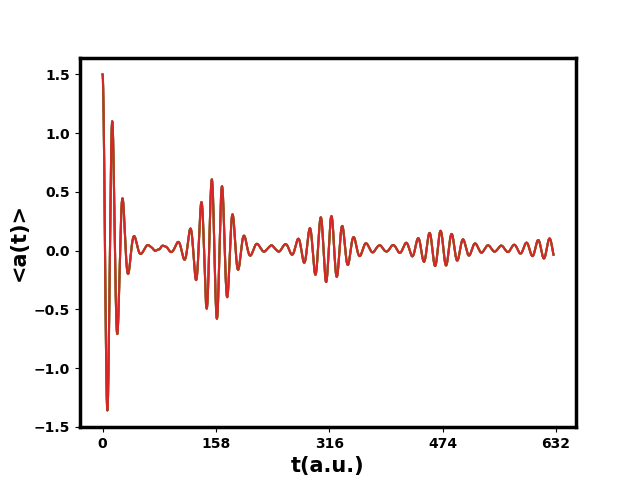}
\caption{$T= 0.3 $K }
\label{fig:7(d)}
\end{subfigure}
\begin{subfigure}{0.45\textwidth}
\includegraphics[height=5cm,width=8.5cm]{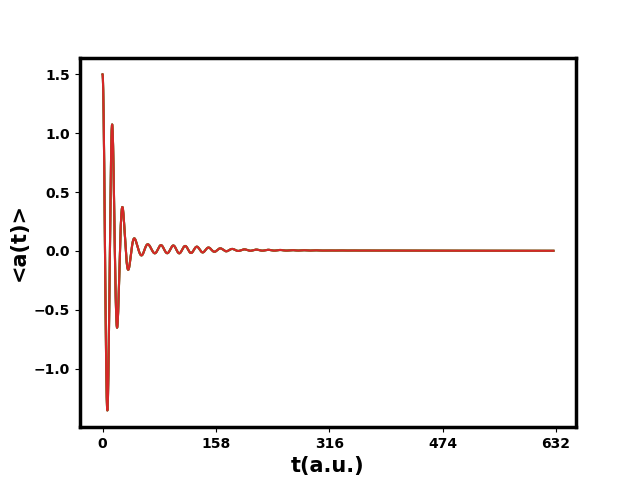}
\caption{$T= 3$ K }
\label{fig:7(e)}
\end{subfigure}
\caption{Effect of bath temperature on quantum state
stored inside optomechanical oscillator with non linearities
$k_c=k_m=0.01$~a.u.  and dissipation
$\gamma_c=\gamma_m=0.00001$~a.u.. The amplitude of revived
states at their revival time are recorded for various values
of bath temperature $T$. }
\label{fig:Figure 7}
\end{figure*}
%%%%%%%%%%%%%%%%
\subsubsection{The effect of non-linearity in system}
%%%%%%%%%%%%%%%%%
Next we study the effect of strong and weak
non-linearity on the revival behavior of the quantum state.
In this case, the dissipation
parameters $\gamma_c$ and $\gamma_m$  for optical and
mechanical modes are chosen to be $0.00001$~a.u.  and
magnitude of Kerr non-linearity  $k_c$ of the optical mode
as well as quadratic anharmonicity  $k_m$ of the mechanical
mode is varied.
For Figure~\ref{fig:6(a)}  the value of the non-linearity
parameter is chosen to be large ( $k_c=k_m = 0.5$~a.u.).
The value of revival time for this non-linearity as calculated from
equation~(\ref{eq:7}) is $6.4$~a.u.   and  indicates that
revivals should appear at these times.  However,
from Figure~\ref{fig:6(a)} one can notice that,  the revival of
quantum state is absent at this value of non-linearity and the pattern is
irregular.  Similar behavior is observed for non-linearity values of $0.05$~a.u.  
(Figure~\ref{fig:6(b)}). Hence, we conclude that high values of
non-linearity do not support the phenomenon of collapse and
revival of quantum state.  For a comparatively
smaller value of non-linearity, $0.005$~a.u.
(Figure~\ref{fig:6(c)}), the constructive and destructive
interference among multi-component Schr\"odinger cat states
makes the collapse and revival to
occur at fixed time without irregularities. Further
decrease in non-linearity value by a factor of 10
(Figure~\ref{fig:6(d)})
decreases the successive amplitude of revived states. This
happens due to the large value of revival time i.e. $6400$
a.u. which slows down the  process of collapse and revival.
As a consequence the interaction time with the dissipative
environment increases leading to substantial fall in
amplitude.
Hence, we conclude that information can be successfully
retrieved when the non-linearities in both the modes is
of the order of $10^{-3}$~a.u..

\begin{figure*}[ht] 
\begin{subfigure}{0.45\textwidth}
\includegraphics[height=5cm,width=8.5cm]{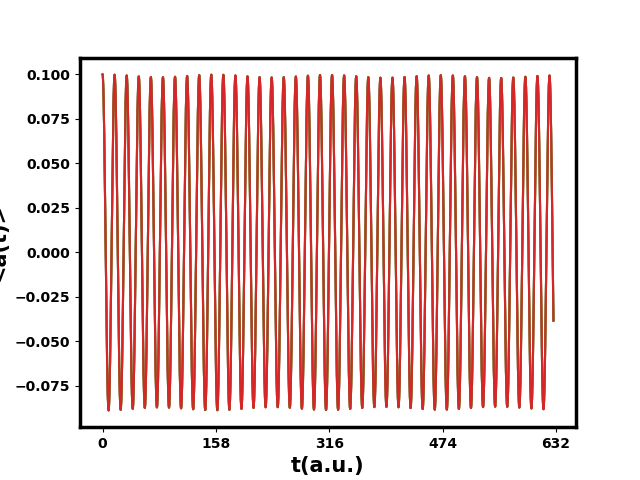}
\caption{Initial amplitude$=0.1$}
\label{fig: 8(a)}
\end{subfigure}
\begin{subfigure}{0.45\textwidth}
\includegraphics[height=5cm,width=8.5cm]{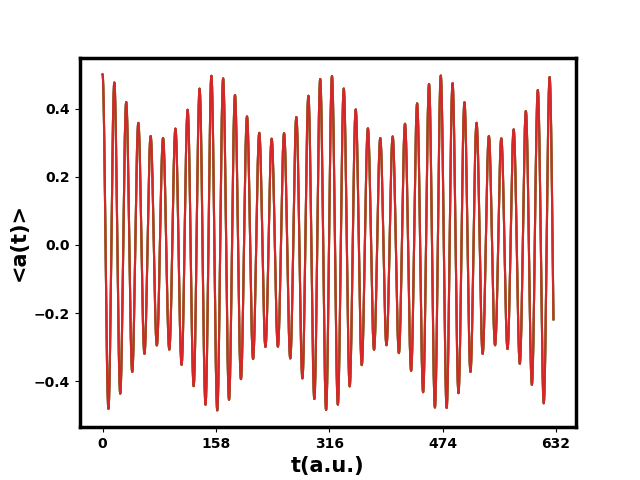}
\caption{Initial amplitude$=0.5$}
\label{fig: 8(b)}
\end{subfigure}
\begin{subfigure}{0.45\textwidth}
\includegraphics[height=5cm,width=8.5cm]{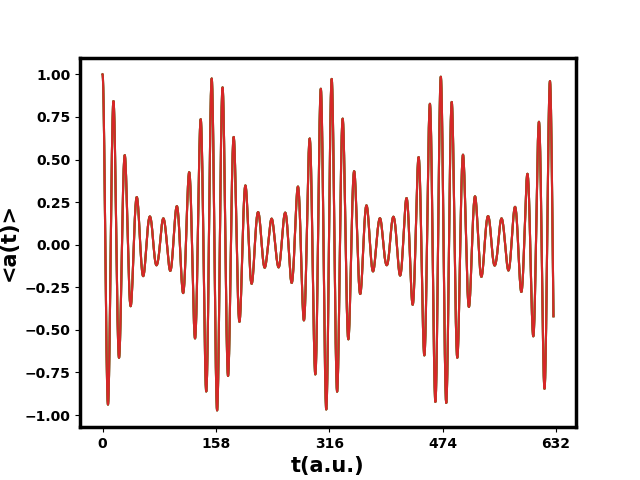}
\caption{Initial amplitude$=1.0$}
\label{fig: 8(c)}
\end{subfigure}
\begin{subfigure}{0.45\textwidth}
\includegraphics[height=5cm,width=8.5cm]{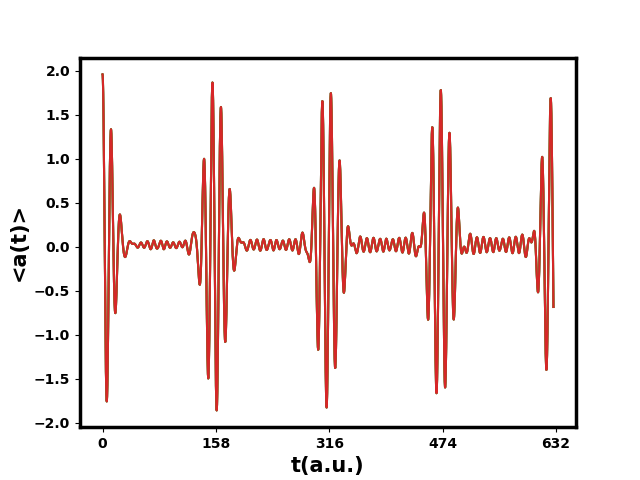}
\caption{Initial amplitude$=2.0$}
\label{fig: 8(d)}
\end{subfigure}
%\begin{subfigure}{0.45\textwidth}
%\includegraphics[height=5cm,width=8.5cm]{fig4.png}
%\caption{Amplitude$=2.0$}
%\label{fig: 8(e)}
%\end{subfigure}
\caption{Collapse and revival pattern of $\langle a(t)
\rangle$ of quantum state being plotted as a function of
time with non-linearities ($k_c$ and $k_m$)  and dissipation
($\gamma_c$ and $\gamma_m$) of optical and mechanical modes
equal to $0.01$~a.u.  and $0.00001$~a.u., respectively.   The
pattern is shown for states with initial amplitude$= 0.1,
0.5, 1.0$ and $2.0$.}
\label{fig:Figure 8}
\end{figure*}
\subsubsection{The effect of bath temperature }
The fact that collapse and revival of a quantum state
show dependence on bath temperature (Equations(\ref{eq:11})
and (\ref{eq:15})),  provides an opportunity to explore the
revival pattern for various values of $T$.  The results of this
study are shown in Figures~\ref{fig:7(b)} -
~\ref{fig:7(e)}.  
Temperatures in $\mu$K and mK range are insignificant
for optical mode, however, these temperature range are
notable for mechanical mode of the optomechanical system.  
%As a result at very small temperatures insignificant effect
%on the collapse revival pattern arising from its optical
%mode but the mechanical mode contributes to the overall
%change in revivals due to variation in the number of
%phonons involved in the interaction. 
At these very small temperatures effect on the collapse and
revival pattern arises from mechanical mode due to variation
in the number of phonons involved in the interaction.  As
shown in Figure~\ref{fig:7(b)} at $\mu$K range the system
behaves in similar manner as it does at 0K (Figure~\ref{fig:Figure 4}).  At mK range
(Figure~\ref{fig:7(c)})  the revival amplitude starts
decreasing with subsequent revival period. This successive
decrease in revival amplitude is due to the increase in mean
thermal phonon number of the bath which creates hindrance
for complete revival in amplitude while interacting with the
system.  Further increase in the value of temperature to 0.3
K  as indicated in Figure~\ref{fig:7(d)} shows exceedingly
large dissipation of the amplitude of quantum state.  At 3
K, collapses and revivals in the quantum state disappear
entirely  owing to noticeable effect of the bath
temperature.
%Amplitude of the coherent state reaches zero  value in few
%moments after it starts evolving and advances with entirely
%collapsed state. n$^{th}_m$ acquires the value of
%approximately equal to 1000 at 3 K. 
%Therefore, the temperature of the bath at 0 K and in $\mu$K
%range shows robustness in results for quantum memory
%system.   

\subsubsection{The effect of initial amplitude of coherent
state}
%%%%%%%%%%%
Initial amplitude of coherent state measures the extent to
which the displacement operator displaces the vacuum state
while constructing the  initial coherent state. 
%This quantum state under time evolution suffers collapses
%and revivals at certain periods of time. 
Our findings highlight the effect  of initial amplitude on the
behavior of collapses and revivals in Figures~\ref{fig:
8(a)} -~\ref{fig: 8(d)}. Here the value of non-linearities
in both the modes is $0.01$~a.u.  and dissipation is
$0.00001$
a.u..  The bath temperature is maintained at 0K. 
%Therefore, the mean thermal photon and phonon occupation
%numbers  at this temperature is zero. 
Behavior of the quantum information is studied at various
values of initial amplitude taken as approximately $0.1$,  $0.5$,  $1.0$ and $2.0$.
%Higher values of amplitude could be accommodated by
%increasing the Hilbert space.
For very small values of the amplitude there is no
perceptible effect on the pattern as seen in
Figure~\ref{fig: 8(a)}, where we have chosen  the amplitude
to be $0.1$. For such small amplitude,  the
system-environment interaction and non-linearity does not
effect the quantum state inside the optomechanical system
and its time evolution is  harmonic in nature.  We note that
the revival behavior of the quantum state in this situation
is similar to the revival behavior of a simple harmonic
oscillator without dissipation.  We term such revivals as
perfect revivals which support retrieval of quantum state
without any loss in amplitude.   As the value of amplitude increases to
$0.5$ (Figure~\ref{fig: 8(b)}), the process of collapses
and revivals is seen in its budding state.  Further
surge in the amplitude to $1.0$ (Figure~\ref{fig:
8(c)}) results in an erratic energy spectrum leading to well
distinguished collapse and subsequent revival of the quantum
state.  At amplitude of $1.5$ (Figure~\ref{fig:Figure 4}),  the collapses are pre-eminent.  
%At last for amplitude of $2.0$~a.u. (Figure~\ref{fig: 8(e)}),
With further increase in amplitude,  the  collapses  and revivals are  accompanied by widened
collapsed phase as shown by loss in amplitude of the
quantum state for a longer period of time(Figure~\ref{fig: 8(d)}).  
Thus, the retrieval of the information must be
carried out at that time instant where revivals are
prominent. 
%%%%%%%%%%%%%%%%%%%%%%%%%%%%%%%%%%%%%%%%%%%%%%%%%%%%%%
\section{Conclusion} \label{4}
We considered a quantum memory device comprising of an
optical mode interacting with a mechanical oscillator
with  nonlinearities present  both in the optical as
well the mechanical mode.  The  time degradation of
quantum information stored in this quantum memory device was
studied under a dissipative environment  and with respect to
nonlinearity parameters.  The quantum master equation
approach was used to analyze the dissipative dynamics
of coherent quantum states which were used for information
storage.  The parameters chosen to monitor the time
degradation of the quantum states were the loss of coherence
as well as the loss of amplitude in an originally coherent
quantum state with a well defined amplitude.  The nature of
coherence was analyzed in terms of Wigner function which
yields useful visual insights of progression of  quantum
states through system non-linearities as well as
decoherences due to interaction with the environment.  We
further evaluated the expectation value of amplitude of the
stored quantum information as a function of time to mark the
extent of overall sustainability of information.  The time
evolution of amplitude is studied for a wide range of
parameters.  Effect of the environment,  non-linearity, bath
temperature and initial amplitude on the revival behavior of
the quantum state is determined.  We also presented the best
suited parameters for the quantum state retrieval.  This
work has potential applications in quantum information
processing and long distance communication.
%%%%%%%%%%%%%%%%%%%%%%%%%%%%%%%%%%%%%%%%
%\bibliography{qmemo.bib}

\end{document}